\documentclass{aa} 

\usepackage{graphicx, enumitem, xcolor}   

\usepackage{txfonts}
\usepackage{array}
\usepackage{hyperref}
\usepackage[normalem]{ulem}

\newcommand{\beq}{\begin{equation}}
\newcommand{\eeq}{\end{equation}}
\newcommand{\beqa}{\begin{eqnarray}}
\newcommand{\eeqa}{\end{eqnarray}}
\newcommand{\eqn}[1]{Eq.\,(\ref{#1})}
\newcommand{\fig}[1]{Fig.\,\ref{#1}}
\newcommand{\tab}[1]{Tab.\,\ref{#1}}
\newcommand{\sect}[1]{Sect.\,\ref{#1}}
\newcommand{\appx}[1]{Appendix\,\ref{#1}}

\newcommand{\ith}{$i$th}
\newcommand{\jth}{$j$th}
\newcommand{\dd}{{\rm d}}
\newcommand{\toot}{\rightleftharpoons}

\makeatletter
\renewcommand*\env@matrix[1][c]{\hskip -\arraycolsep
  \let\@ifnextchar\new@ifnextchar
  \array{*\c@MaxMatrixCols #1}}
\makeatother

\def\lsim{\mathrel{\rlap{\lower 3pt \hbox{$\sim$}} \raise 2.0pt \hbox{$<$}}}
\def\gsim{\mathrel{\rlap{\lower 3pt \hbox{$\sim$}} \raise 2.0pt \hbox{$>$}}}

\begin{document}

\title{Reducing the complexity of chemical networks via interpretable autoencoders}

\subtitle{}

\author{
    T. Grassi\inst{1,2,}\thanks{E-mail: tgrassi@usm.lmu.de}
    \and
    F. Nauman\inst{3}
    \and
    J. P. Ramsey\inst{4}
    \and
    S. Bovino\inst{5}
    \and
    G. Picogna\inst{1,2}
    \and
    B. Ercolano\inst{1,2}
}

\institute{
    Universit\"ats-Sternwarte M\"unchen, Scheinerstr. 1, D-81679 M\"unchen, Germany
    \and
    Excellence Cluster Origin and Structure of the Universe, Boltzmannstr.2, D-85748 Garching bei M\"unchen, Germany
    \and
    2MNordic IT Consulting AB, Sk\aa rs led 3, 412 63 Gothenburg, Sweden.
    \and
    Department of Astronomy, University of Virginia, Charlottesville, VA 22904, USA
    \and
    Departamento de Astronom\'ia, Facultad Ciencias F\'isicas y Matem\'aticas, Universidad de Concepci\'on,\\
    Av. Esteban Iturra s/n Barrio Universitario, Casilla 160, Concepci\'on, Chile
}

\date{Accepted XXX. Received YYY; in original form ZZZ}

\abstract{
In many astrophysical applications, the cost of solving a chemical network represented by a system of ordinary differential equations (ODEs) grows significantly with the size of the network, and can often represent a significant computational bottleneck, particularly in coupled chemo-dynamical models. Although standard numerical techniques and complex solutions tailored to thermochemistry can somewhat reduce the cost, more recently, machine learning algorithms have begun to attack this challenge via data-driven dimensional reduction techniques. In this work, we present a new class of methods that take advantage of machine learning techniques to reduce complex data sets (autoencoders), the optimization of multi-parameter systems (standard backpropagation), and the robustness of well-established ODE solvers to to explicitly incorporate time-dependence. This new method allows us to find a compressed and simplified version of a large chemical network in a semi-automated fashion that can be solved with a standard ODE solver, while also enabling interpretability of the compressed, latent network. As a proof of concept, we tested the method on an astrophysically-relevant chemical network with 29~species and 224~reactions, obtaining a reduced but representative network with only 5~species and 12~reactions, and a $\times65$ speed-up.
}


\maketitle


\section{Introduction}
Over the last few decades, the technological advance and proliferation of astronomical observations has revealed a dramatic richness and variety of chemical species in space, in particular in the interstellar medium (ISM), star-forming regions and protoplanetary disks \citep[e.g.][]{Henning2013,Walsh2014}.  Especially at radio, sub-mm, and infra-red wavelengths, sensitivity and spatial resolution improvements have resulted in the discovery of a plethora of both simple and complex chemical species \citep{McGuire2018}. Understanding how these species form, how they react with other species, and what physical conditions they trace requires (both experimental and theoretical) accurate and comprehensive tools  \citep{Herbst2009,Jorgensen2020}.

From a numerical point of view, the time-dependent study of the formation and evolution of key chemical tracers under ISM conditions is obtained by solving a system of ordinary differential equations (ODEs) representing the time evolution of the species interconnected by a set of chemical reactions. Increasing molecular complexity leads to a larger number of species and reactions that must be included in the chemical network to realistically represent a specific physical environment. This is particularly true for the ISM and protoplanetary disks as we learn more about these environments through observations. As such, the size of requisite chemical networks increases quickly, as demonstrated for example by the hundreds of species and the thousands of reactions in the KIDA database of chemical reactions\footnote{\url{http://kida.astrophy.u-bordeaux.fr/}} \citep{Wakelam2012}. Naturally, increasing network complexity leads to increasing computational costs for solving the associated system of ODEs, which indeed becomes prohibitive when coupled to dynamical models that follow the evolution of astrophysical environments over space and time.

As a result, in order to keep costs reasonable, existing chemical models are typically far from being complete. While 1D and 2D astrochemical simulations \citep[e.g.][to name only a few]{Garrod2008,Bruderer2009,Woitke2009,Semenov2010,Sipila2010,Akimkin2013,Rab2017} enable us to deeply explore the chemistry and constrain the sensitivity of chemical networks to various parameters, the effects of dynamics, magnetic fields and complex radiative transfer, are seldom modeled, and can sometimes lead to a misleading interpretation of the results. On the other hand, (radiative) (magneto-)hydrodynamic simulations are often limited by their computational cost, and when chemistry and microphysics are included, they are often done so only with simplified prescriptions. Several efforts have been made in this respect \citep[e.g.][]{Glover2010,Bai2016,Ilee2017,Bovino2019,Grassi2019,Gressel2020,Lupi2020}, but the chemical complexity is intentionally limited to the species that are relevant for the given problem (e.g.~coolants). Analogously, Monte Carlo approaches for constraining reaction rates from observations using Bayesian inference (e.g.~\citealt{Holdship2018}) represent another situation where faster computation of the chemistry would be beneficial.

Employing specific simplified prescriptions is an example of a so-called reduction method that attempts to preserve the impact of the relevant (thermo)chemistry on dynamical evolution while minimizing the cost. As already alluded to, practically, this usually means removing species and reactions based on chemical and physical considerations (e.g.~\citealt{Glover2010}). However, as with any reduction method, this approach is tuned to the environment of interest, and can result in the introduction of uncertainties when the problem becomes non-linear or when the validity of the reduction has not been tested in a particular region of parameter space.
\citet{Grassi2012}, following \citet{Tupper2002}, presented an automatic reaction rate-based reduction scheme, yielding a variable speed-up while also providing good agreement with results from the complete network. Recently, \citet{Yoon2018} and \citet{Xu2019} employed similar approaches for the study of molecular clouds and protoplanetary disks. Meanwhile, \citet{Ruffle2002,Wiebe2003,Semenov2004} compared the reactions kinetic with a method similar to the so-called ``objective reduction techniques'' from combustion chemistry. Conversely, \citet{Grassi2013} proposed a topological method to reduce the complexity of a chemical network, i.e.~``hub'' chemical species are selected for a reduced network after being predicted to be more active/important during the evolution of the system. Yet another approach is to split the reaction time-scales to reduce the computational complexity of the problem, for example by integrating the slow scales only and add the fast ones as a correction term (e.g.~\citealt{Valorani2001,Nicolini2013}). The methods described above can provide non-negligible speed-ups relative to the cost of computing the complete network, but they can be problem-dependent, difficult to implement in practice (e.g. to couple to a dynamical model), or interfere with other reduction techniques and lose some of their effectiveness when included in large-scale dynamical models. 

An alternative to these methods is to employ data-driven machine learning techniques \citep{Grassi2011,deMijolla2019}. The evolution of a standard chemical solver can be replaced by a more computationally-tractable operator that is capable of advancing the solution in time. In contrast to a physically motivated approach (i.e.~removing reactions and species based on the environment or problem), here the chemical and physical assumptions are minimal, the simplifications automated, but the interpretation is often much more difficult \citep[e.g.][]{Chakraborty2017}.

More recently, it has been demonstrated that machine learning algorithms can also be used to identify so-called ``governing'' equations \citep{Brunton2016,Long2017,Chen2018,Rackauckas2019,Raissi2018,Raissi2019,Choudhary2020}, solve forward and inverse differential equation problems efficiently \citep{Rubanova2019}, and construct low dimensional interpretable representations of physical systems \citep{Champion2019,Wiewel2018,Yildiz2019}. Excluding \citet{Hoffmann2019}, where an effective reaction network is inferred from observations with a sparse tensor regression method, these methods have never been applied to the complicated system of coupled ordinary differential equations that is commonly used to represent chemical evolution.

In this work, we present a new deep machine learning method that can discover a low-dimensional chemical network via autoencoders and that effectively represents the dynamics of a full network. The two main advantages of this approach are that (i) it is more interpretable compared to existing data-driven approaches and (ii) the resulting low-dimensional ODE system can be easily integrated with standard ODE solvers and coupled to hydrodynamic simulations where the calculation of time-dependent chemistry is desired. We also present a proof-of-concept application of the method.

The layout of this work is as follows. In \sect{sect:methods} we review the application of ODEs to chemistry, other reduction techniques, and how we apply autoencoders to discover a low-dimensional network. In \sect{sect:results} we apply our method to an astrophysical chemical network and compare the results to a calculation with a full network. We discuss the limitations and the potential solutions in \sect{sect:limitations}. Conclusions are in \sect{sect:conclusions}.

\section{Methods}\label{sect:methods}
\subsection{Systems of chemical ordinary differential equations}\label{sect:ode}

The time evolution of the species in a chemical network is commonly represented by a set of coupled ODEs defined by the species and reactions in that network. The definition of each reaction includes a rate coefficient, $k$, that represents the probability of the reactants to form the products. The variation in time of the abundances of each species, $n_i(t)$, is given by
\beq\label{eqn:ode}
 \frac{\dd n_i}{\dd t} \equiv \dot n_i = - n_i\sum_{j}k_{ij} n_j + \sum_{j} n_j \sum_{l}  k_{jl} n_l\,,
\eeq
where the first term on the right-hand side represents the reactions involving the destruction of the \ith{} species, while the second its formation, and the sums are over $N$ total species. In \eqn{eqn:ode}, the abundances are functions of time, and $k$ can depend on temperature, and in some environments density, and that may both vary with time \citep[e.g.][]{Baulch2005}. If $\bar x$ represents the abundances of the $N$ species, \eqn{eqn:ode} can be written in the more general form 
\beq\label{eqn:fode}
  \dot{\bar x} = f(\bar x; k)\,,
\eeq
i.e., as an operator $f$ in $\mathbb{R}^N\to\mathbb{R}^N$, defined by the rate coefficients $k$ and the input variables $\bar x(t)$. In many astrophysical applications, the number of species $N$, and hence the number of differential equations in the system, ranges from a few tens to several hundreds, while the number of reactions subsequently range from hundreds to tens of thousands reactions depending on the complexity of the astrophysical model (see \appx{sect:relation}). In general, numerically solving a system of ODEs for a large number of species and reactions requires a non-negligible amount of computational resources, and usually represents a numerical bottleneck for many applications, particularly for those that couple chemistry to dynamics. The two main reasons for this are (i) the substantial coupling between different species/ordinary differential equations and (ii) the interplay between the desired solution accuracy and different time scales in the problem that results in small integration (time) steps \citep[e.g.][]{Bovino2013,Grassi2013}. Both effects, however, can result in a so-called ``stiff'' system of ODEs, which further increases the challenges by requiring (generally) more expensive ODE solvers with particular features.

\subsection{Reduction techniques}\label{sect:reduction}

The simplest approach to reducing the cost of computing chemistry is to exploit available computer hardware (e.g.~vectorization on modern central processing units [CPUs], graphics processing units [GPUs], or performance analysis and tuning), which can lead to a moderate speed-up. However, vectorization is practically limited by how effectively mathematical operators can be computed in parallel across array elements \citep[e.g.][]{Tian2013}. GPU-based methods are meanwhile limited by the technical specifications of the hardware \citep{Curtis2017}. Certainly, these approaches reduce the cost, but they are still far from eliminating the numerical bottleneck that is computational chemistry.

Another approach is to make linear fits to complex reaction rate coefficients that may depend on exponential or logarithmic functions. This reduces the cost of evaluating $f(\bar x;k)$ and the Jacobian ($\partial f_i/\partial x_j$), but does not remove the complexity of the problem itself, and therefore only achieves a relevant, but relatively moderate, speed-up.

Among other more generic solutions, some ODE solvers can take advantage of the sparsity of Jacobian matrices, a distinguishing feature of the chemical ODE systems. Applying a compression algorithm to the Jacobian can reduce the computational cost considerably \citep{Duff1986,Hindmarsh2005,Nejad2005,Perini2012}, but the time required to solve the chemical network over time in several astrophysical applications remains still non-negligible.

More commonly however, both in astrophysics and in chemical engineering, the computational impact of the chemistry is reduced by determining which reactions are relevant for a given problem, for example via expert inspection \citep[e.g.][]{Glover2012,Gong2017}, or via rate efficiencies evaluation \citep[e.g][]{Grassi2012,Xu2019}. However, these methods may fail to reproduce the detailed evolution of less abundant species, or may become ineffective when the network cannot be substantially reduced (e.g.~because the reactions contribute to the stiffness of the system).

In more recent years, machine learning and Bayesian methods are beginning to be exploited, in particular because of the emergence of a number of tools that allow these techniques to be implemented with relative ease (e.g.~\citealt{Grassi2011,deMijolla2019,Heyl2020}). In particular, (deep) neural networks have been employed to predict the evolution in time of the chemical abundances and temperature (recently termed ``emulators''), and replace the ODE solver within the parameter space described by the supplied training data. This approach is certainly one of the most promising to negate the computational cost of solving chemistry, but it has so far been limited by complicated (and not necessary successful) neural network training sessions, error propagation, and a lack of interpretability. Interpretability, i.e.~have a certain degree of machine learning model's transparency \citep{Lipton2016,Miller2017}, can become relevant when the input conditions of the neural network lie outside the training, test, and/or validation sets, or if the training is affected by unnoticed under-/over-fitting to the data. This limit to interpretability resides in the intrinsic design of deep neural networks (DNNs), i.e.~thousands of parameters that define the interaction of non-linear functions. For comparison, analogous machine learning techniques, such as principal component analysis, have some degree of interpretability \citep{Shlens2014}, while DNNs applied to chemistry apparently do not \citep{Chakraborty2017}. Despite these drawbacks, given the constant improvement of machine learning techniques and hardware, deep machine learning techniques are one of the most promising reduction methods for chemistry proposed in the last several years.

The technique presented in this paper is somewhat in between the methods described above. We want to take advantage of the capability of DNNs to find simplified representations of complex chemical networks, but coupled with the time-dependent accuracy provided by ODE solvers, while also including the interpretability of the resulting network. 

\begin{figure}[tb]
  \includegraphics[width=\linewidth]{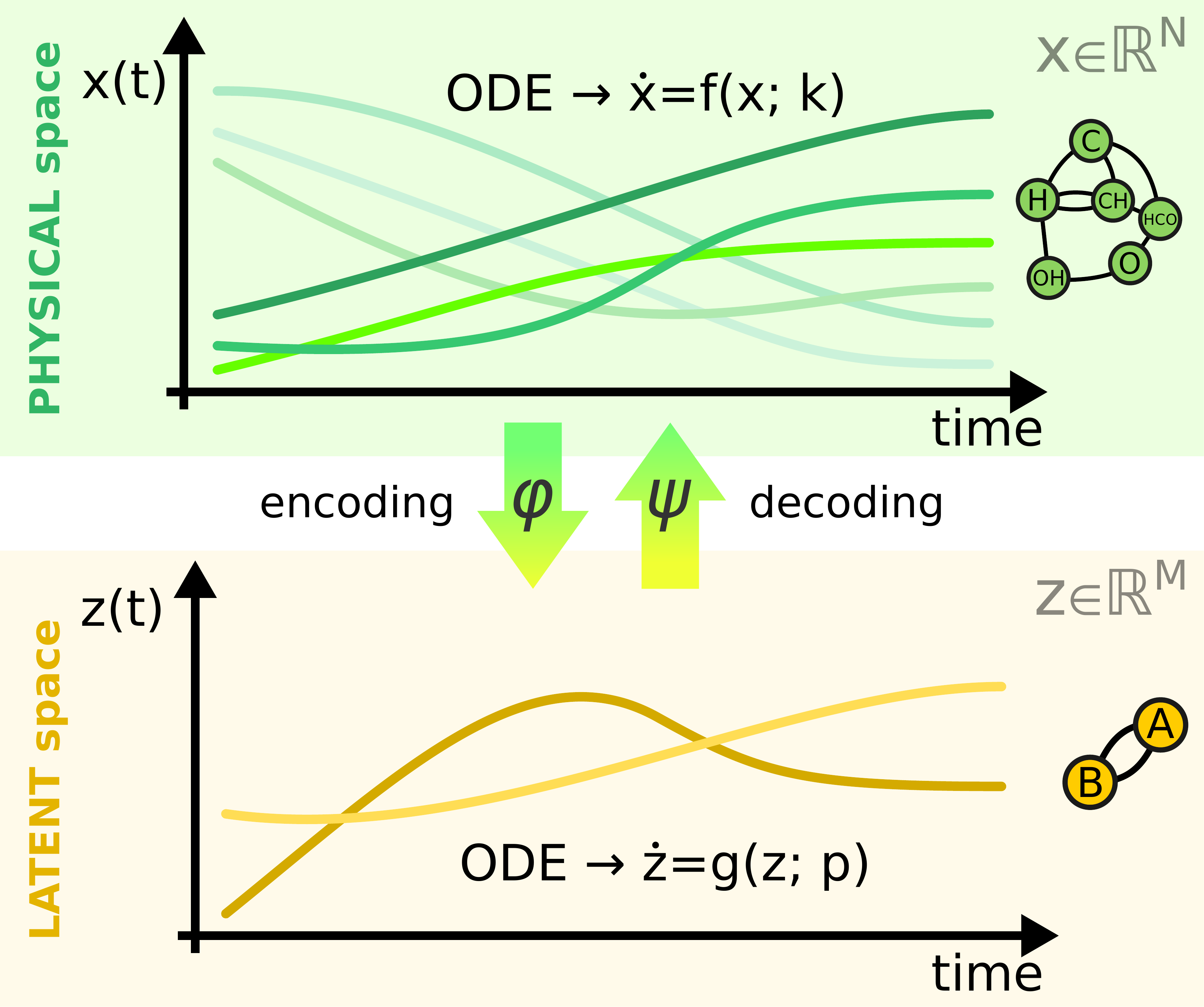}
  \caption{Graphical representation of the proposed method. The evolution of the chemical species $\bar x(t)$ in the physical space is usually obtained by integrating $f(\bar x; k)$ in time (upper panel). This can be reconstructed by evolving a different set of variables in the latent space $\bar z(t)$ by using the $g(\bar z; p)$ (lower panel). The transformation between the two spaces is obtained with an encoder ($\varphi$) and a decoder ($\psi$). The physical space has $N$ chemical species/dimensions ($N=6$ in this schematic), while the latent space has $M<N$ variables/dimensions ($M=2$ in this schematic). Lines schematically show how the abundances in the physical and latent spaces might evolve with time. Sketches of the chemical networks in the physical space and the latent space are shown on the right side. The latent network (A$\toot$B) has less species and reactions, but captures the dynamics of the full chemical network faithfully.}\label{fig:method}
\end{figure}

To accomplish this, we reduce the dimensionality of the physical space with an encoder ($\varphi$) in order to create the so-called latent space (see \fig{fig:method}). In the latent space, the $N$ abundances of species, $\bar x$, are represented by another set of $M$ ($<N$) abundances: $\bar z$. We then postulate that these variables belong to another chemical network defined in the latent space. Analogously to the chemical network in the physical space, i.e., that is represented by the system of ODEs $f(\bar x; k)$ in \eqn{eqn:ode}, the chemical network in the latent space still evolves following a set of differential equations (albeit a different set), $g(\bar z; p)$, where the parameters $p$ play the same role as the rate coefficients $k$ do in the physical space. Using $g$, it is possible to evolve $\bar z(t)$ forward in time in the latent space. Next, a decoder ($\psi$) transforms the variables, $\bar z(t)$, back to the physical abundances, $\bar x(t)$. Our method aims at finding the encoder $\varphi$, decoder $\psi$, and the operator $g$, which then allows us to obtain $\bar x(t)$ from the evolving $\bar z(t)$, which has a significantly smaller number of dimensions and hence is much less computationally expensive to integrate.

\subsection{Autoencoders}\label{sect:autoencoders}

Autoencoders are a widely used machine learning techniques in many disciplines that incorporate symmetric pairs of deep neural networks, and have applications in denoising images \citep[e.g.][]{Gondara2016}, generating original data with variational autoencoders \citep[e.g.][]{Kingma2013}, detecting anomalies in images and time-series \citep[e.g.][]{Zhou2017}, and, more relevant to this work, to reduce the dimensionality of data \citep[e.g.][]{Kramer1991}. Autoencoders are a non-linear generalization of principal component analysis \citep{Jolliffe2002} with trainable parameters, which makes them suitable for all kinds of dimensional reduction tasks, including reducing the complexity of chemical networks.

\begin{figure}
 \includegraphics[width=\linewidth]{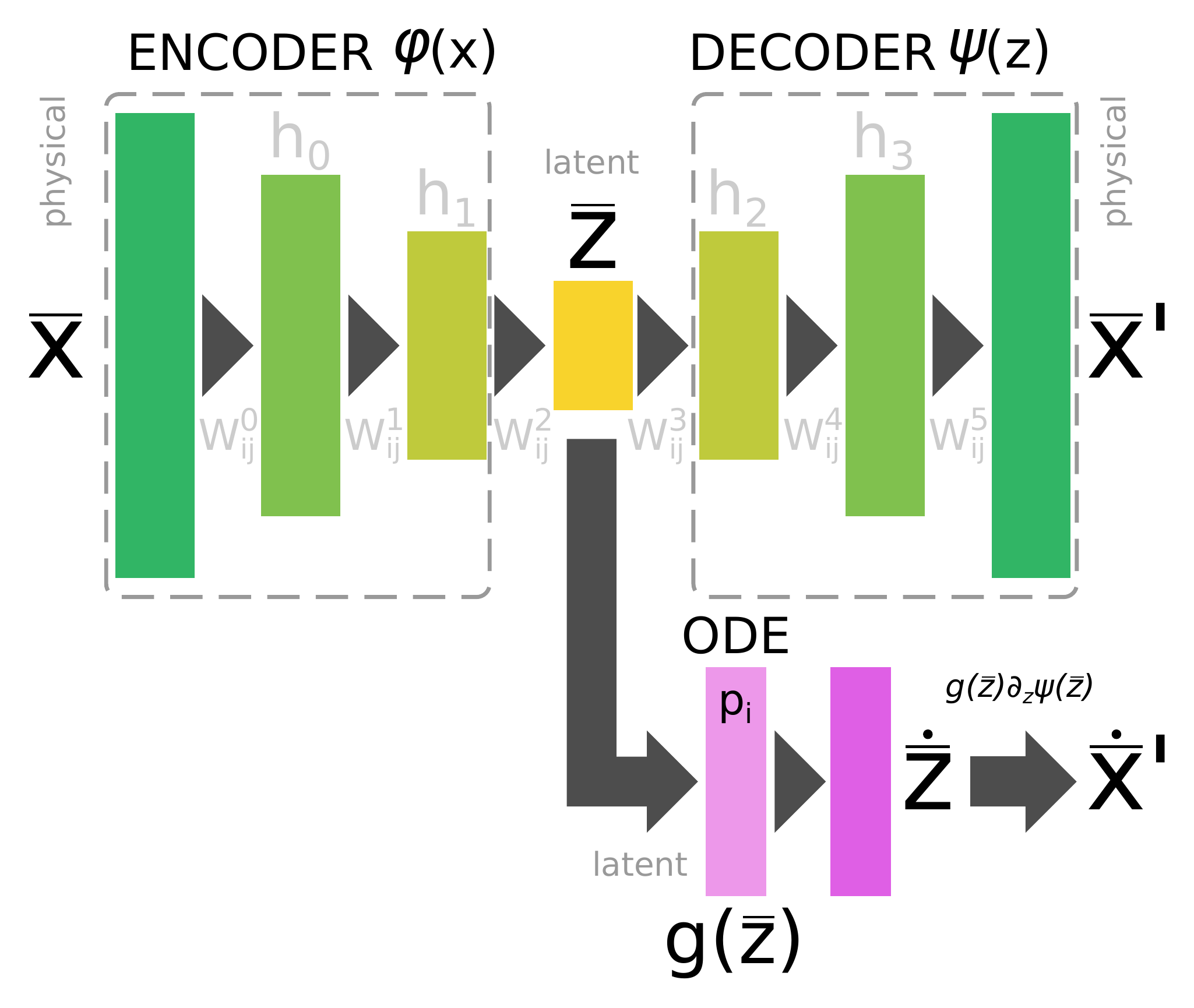}
 \caption{Schematic of the autoencoder and latent ODE system. The upper part of the sketch represents the autoencoder, with both encoder and decoder deep neural networks. Each rectangle represents a layer of the deep neural network, linked together by weights $W$ and biases $b$ (omitted for the sake of clarity). The input to the encoder is $\bar x$, with $N$ nodes/dimensions, connected to a sequence of hidden layers $h_i$ with decreasing dimensionality/number of nodes, until reaching the layer $\bar z$ with $M$ nodes/dimensions, where the maximum compression is obtained. The decoder is symmetric w.r.t.\ the encoder, with layers of increasing dimensionality, ending with an output layer of $N$ nodes/dimensions. Note that, in our case, we have $6$~hidden layers instead of the $4$~shown in this sketch. In the lower part of the sketch, we show the latent ODE system that uses $\bar z$ as inputs and produces $\dot{\bar z}$ as output, both with $M$ dimensions. This additional neural network is controlled by the parameters $p$ (one for each latent reaction), and has an analytical representation $g(\bar z; p)$. The obtained latent space derivatives are decoded to the target derivatives $\dot{\bar x}$ with the same procedure as of \eqn{eqn:loss2}.}\label{fig:sketch}
\end{figure}

Given a set of $N$-dimensional data, $\bar x\in\mathbb{R}^N$, an autoencoder consists of two operators, namely an encoder ($\varphi$ as $\mathbb{R}^N\to\mathbb{R}^M$) and a decoder ($\psi$ as $\mathbb{R}^M\to\mathbb{R}^N$), which are optimized to have $\bar x \simeq \bar x' = \psi(\varphi(\bar x))$, i.e., the autoencoder should be capable of reproducing the input data within a given accuracy. The first operator produces the encoded data $\bar z = \varphi(\bar x)$, where $\bar z\in\mathbb{R}^M$, with $M<N$ (upper part of \fig{fig:sketch}). The second operator then reconstructs/decodes $\bar z$ to retrieve $\bar x' = \psi(\bar z)$. The space where $\bar x$ resides is called the \emph{physical space}, while $\bar z$ belongs to the \emph{latent space}. The latent space has a lower dimensionality relative to the physical space. The reconstruction error in $\bar x'$ is the $\ell^2$-norm between the original and the reconstructed data
\beq\label{eqn:loss0}
  L_0 = ||\bar x - \bar x'||^2_2\,,
\eeq
where $L_0=0$ represents perfect reconstruction. If $\bar x$ has been reconstructed properly (i.e.~$\bar x = \bar x'$), the amount of information is conserved through the autoencoder, and hence the information contained in $\bar z$ is sufficient to represent $\bar x$, but with the advantage of a lower dimensionality. In \fig{fig:sketch} (upper part), we schematically show the autoencoder as a set of layers of a deep neural network. The first layer on the left is the input layer with $N$ nodes (also called neurons), which is connected to a second hidden layer, $\bar h_0$, by weights $W_{ij}^0$ and biases $b^0_j$ (not shown) via $\bar h_0={\cal F}_{\rm a}(W\times \bar x + \bar b)$, where ${\cal F}_{\rm a}$ is a so-called activation function. We add hidden layers, each with less nodes than the previous layer, until the layer marked with $\bar z$ is reached. This layer has $M$ nodes and is the representation of $\bar x$ in the latent space. The decoder works in an analogous way, connecting $\bar z$ to $\bar x'$ via several hidden layers with weights and biases between the layers, with each layer having a larger number of nodes until reaching the last, output layer that consists of $N$ nodes (i.e.~the same number of nodes as the input layer). The aim of the neural network training is then to find the optimal values for the weights $W_{ij}^\ell$ and the biases $b^\ell_j$ that minimize the loss $L_0$ by using a gradient descent technique and an optimizer method that computes the adaptive learning rates for each parameter \citep{Rumelhart1988,Lecun1998}.

In a chemical network, $\bar x$ represents the abundances of the $N$ chemical species that compose the network. In a successfully trained autoencoder, $\bar z = \varphi(\bar x)$ represents the $N$ chemical abundances $\bar x$ in the latent space with $M<N$ variables. In other words, the latent space contains the \emph{same} information as the physical space, but in a compressed format with less variables. However, with only a pure autoencoder, some important information (in particular, the time derivatives of the abundances) cannot easily be inferred, and the latent space has no obvious chemical or physical interpretation. For this reason, we need to extend the deep neural network autoencoder with an additional ``branch''.

\subsection{Differential equation identification in latent space via autoencoders}\label{sect:autoencoder_plus}

To enable the interpretability of the encoded data we need to have access not only to $\bar z$, but also to the time derivative $\dot{\bar z}$.
We follow an approach similar to \citet{Champion2019}, that couples autoencoders with a ``sparse identification of nonlinear dynamics'' algorithm \citep[\textsc{SINDy};][]{Brunton2016}. \textsc{SINDy} consists of a library of (non-)linear functions (e.g.~constant, polynomial, and trigonometric) whose activation is controlled by a set of parameters that determine the importance of each term (i.e.~weights), and is intended to represent the right-hand side of a generic non-linear system of ODEs. Rather than using the \textsc{SINDy} algorithm directly, we instead use the time derivatives as additional constraints and employ a set of functions that mimic the right-hand side of a chemical ODE system during the training phase.

The time derivative\footnote{We define $\partial_t \equiv \partial/\partial t$, $\partial_x\equiv\partial/\partial x$, and $\partial_z\equiv\partial/\partial z$.} of the latent variables, $\partial_t{\bar z} = \partial_t{\varphi(\bar x)}$, after a change of variables, can be written as $\dot{\bar x}\,\partial_x\varphi(\bar x)$. Since we are seeking a ``compressed'' chemical network in the latent space, we also define $\dot{\bar z} = g(\bar z; p)$ as analogous to \eqn{eqn:fode}, where $p$ are the unknown latent rate coefficients. This allows us to define an additional $\ell^2$-norm loss,
\beq\label{eqn:loss1}
  L_1 = ||g(\bar z; p)- \dot{\bar x}\,\partial_x\varphi(\bar x)||^2_2\,,
\eeq
that we can minimize on during training.

Analogously, $\dot{\bar x}$ can be written as $\dot{\bar z}\,\partial_z\psi(\bar z)$, and we can define the corresponding loss term
\beq\label{eqn:loss2}
  L_2 = ||\dot{\bar x} - g(\bar z; p)\,\partial_z\psi(\bar z)||^2_2\,,
\eeq
where $\dot{\bar x}$ is known from the chemical evolution in the physical space. $L_1$ and $L_2$ are the losses that control the reconstruction of $\dot{\bar x}$ and $\dot{\bar z}$, by constraining $\varphi$, $\psi$, and $g$ at the same time.

To enable the calculation and use of $L_1$ and $L_2$ in our set-up, we include an additional branch in our neural network framework (see lower part of \fig{fig:sketch}) that consists of an input layer taking $\bar z$, produces output $\dot {\bar z} = g(\bar z; p)$, and includes trainable parameters $p$. Defining an analytical form for $g(\bar z; p)$ requires some knowledge of the evolution of $\bar z$ in the latent space. \citet{Champion2019} employ a set of polynomials and arbitrary complicated functions through the \textsc{SINDy} method that are controlled (i.e.~turned on or off) by the weights $\Xi$, an additional $\ell^1$-norm loss term $||\Xi||_1$, and a ``selective thresholding'' approach to minimize the number of functions (a concept referred to as ``parsimonious models''; \citealt{Tibshirani1996,Champion2019b}).

In our case, we choose a form for $g(\bar z; p)$ that represents a chemical network analogous to \eqn{eqn:fode}, but in the latent space, and has the advantage of being interpretable because it is similar to the physical space representation. Furthermore, a natural constraint to place on $g(\bar z; p)$ to ensure its connection to the physical representation is the mass conservation in the latent space, $\sum_i m_i\,z_i=$\,constant, where $m_i$ is the mass of the latent species $z_i$ (see \appx{appx:latent_network}). Rather than defining a specific total mass, we constraint the system only to have constant mass, i.e., $\partial_t \sum_i m_i\,z_i = \sum_i m_i \,\partial_t z_i =0$. This criteria can also be described by a loss function:
\beq\label{eqn:loss_mass}
    L_3 = \left|\sum_{i=0}^{M-1} m_i \left[\dot x\,\partial_x\varphi(\bar x)\right]_i \right| + \left|\sum_{i=0}^{M-1} m_i\, g(\bar z; p)_i\right|\,,
\eeq
where the two terms are the mass conservation criteria according to the encoder and the latent ODE system, respectively.

The remaining task is to design $g(\bar z; p)$ to represent a chemical network. In principle, one could take a brute-force method and include all possible combinations of chemical reactions among the different species. This approach might work for a few latent dimensions, but if we want to include all the connections, when the number of latent variables becomes large, the latent chemical network might become larger than the network in the physical space. In this work, we intentionally limit ourselves to a small number of latent dimensions ($M=5$; see \ref{sect:implementation}), which allows us to construct a small latent chemical network even when including all the possible reactant combinations and requiring mass conservation.  We report our latent chemical network in \appx{appx:latent_network}. We do not currently include any other constraints (as e.g.~parsimonious representations), but we are working to improve the design of $g(\bar z; p)$ in a forthcoming work (see also \sect{sect:limitations}).

Finally, we define the total loss
\beq\label{eqn:loss_tot}
    L = L_0 + \lambda_1 L_1 + \lambda_2 L_2 + \lambda_3 L_3\,,
\eeq
where $\lambda_1$, $\lambda_2$, and $\lambda_3$ are adjustable weights to scale each loss in order to have comparable magnitudes when summed together (discussed more in detail in \sect{sect:limitations}). The loss is minimized using $\bar x$ and $\dot{\bar x}$ from a set of chemical models evolved in time with a standard ODE chemical solver as training data. They represent the ground truth. The aim of the training is to find the parameters of the encoder $\varphi$, the decoder $\psi$, and the latent ODE system $g$ in order to obtain $|L|<\varepsilon$, where $\varepsilon$ is a threshold determined by the absolute tolerance required by the specific astrophysical problem. However, according to our tests, this condition alone does not always guarantee a satisfactory reconstruction of $\bar x(t)$ from the evolved $\bar z(t)$, and, in practice, for the purposes of this first study, we halt the training when the trajectories of the species in the test set are reasonably reproduced by visual inspection (see also \sect{sect:limitations}).

\begin{figure}
 \includegraphics[width=\linewidth]{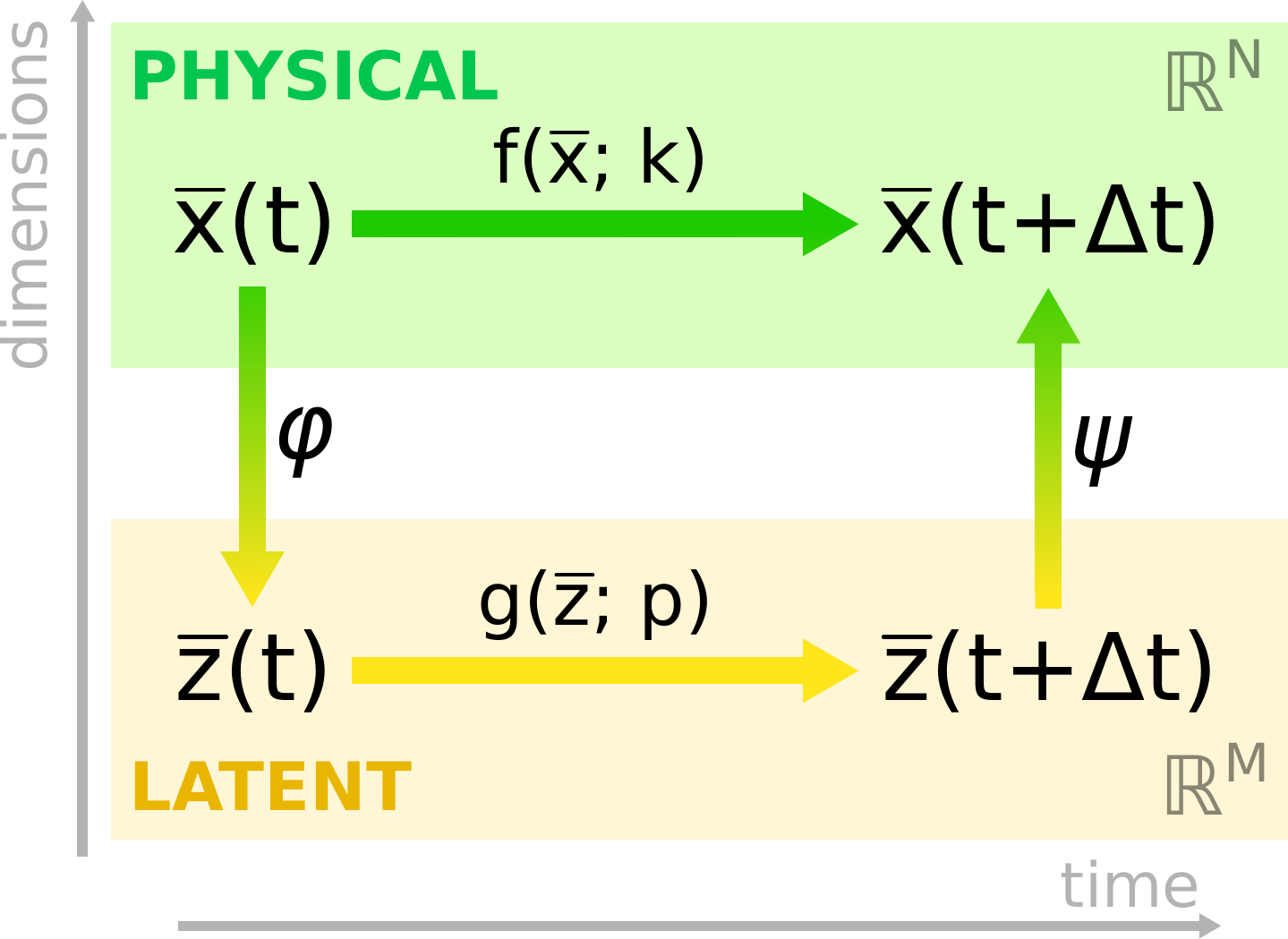}
 \caption{Schematic of the relation between the operators $f$, $g$, $\varphi$, and $\psi$ employed in the current set-up. The ODE system $f(\bar x; k)$, $\mathbb{R}^N\to\mathbb{R}^N$, allows the integration of $\bar x(t)\in \mathbb{R}^N$ in time, and is analogous to $g(\bar z; p)$, $\mathbb{R}^M\to\mathbb{R}^M$ that is employed to integrate $\bar z(t)\in \mathbb{R}^M$. To transform $\bar x$ to $\bar z$ (and vice versa) we use an encoder $\varphi$, $\mathbb{R}^N\to\mathbb{R}^M$, and a decoder $\psi$, $\mathbb{R}^M\to\mathbb{R}^N$. Note that to evolve both $\bar x$ and $\bar z$ we employ a standard implicit ODE solver, hence the time-step $\Delta t$ can be arbitrarily long.}\label{fig:recap}
\end{figure}

To recap (cfr.~\fig{fig:recap}), the time evolution of $\bar x(t)\in\mathbb{R}^N$ that is normally obtained with an ODE solver by integrating $f(\bar x; k)$, can be found by instead evolving $\bar z(t)\in\mathbb{R}^M$ using an ODE solver but a simpler system of ODEs, $g(\bar z; p)$, in a latent space. The encoder $\varphi$ ($\mathbb{R}^N\to\mathbb{R}^M$) and the decoder $\psi$ ($\mathbb{R}^M\to\mathbb{R}^N$) transform $\bar x(t)$ into $\bar z(t)$, and vice versa. The advantage of this method is the explicit inclusion of time-dependence and an ODE solver that integrates a system of equations that is simpler than the original ($M\ll N$), resulting in a method that is not only faster, but also retains some physical interpretability.

\subsection{Implementation}\label{sect:implementation}

Designing a neural network requires determining the nominal number of layers, the interaction among the nodes, choosing an appropriate loss function, and many other aspects. The optimal choice for these hyperparameters can only be determined by training the neural networks on the data many times. The training process optimizes the weights and biases of the network along with any other trainable parameters, for example, parameters $p$ in $g$ in our case. Below, we describe the steps of our implementation, including those that consist of design choices, hyperparameters and trainable parameters (caveats are discussed in \sect{sect:limitations}):
\begin{itemize}
 \item Prepare the training set; calculate the temporal evolution of $\bar x(t)$ and $\dot{\bar x}(t)$ and vary the initial conditions to cover the physical parameter space relevant for the current problem.
 \item Define $M$, the number of dimensions in the latent space.
 \item Design the latent chemical network giving the analytical form of $g(\bar z;p)$; see \sect{sect:autoencoder_plus}. The analytical form of $g$ needs to be provided, but the parameters $p$ are optimized as part of the training.
 \item Determine the importance of each loss weight $\lambda$ in \eqn{eqn:loss_tot}, normalizing each loss to contribute to the total loss.
 \item Train the deep neural network in \fig{fig:sketch} based on minization of the total loss $L$ given by \eqn{eqn:loss_tot}.
\end{itemize}

After a successful training, it is possible to compress the initial conditions $\bar x(t=0)$ defined in the physical space into the latent space variables $\bar z(t=0)$, via the encoder $\varphi$. Therein, $\bar z(t)$ is advanced in time by using a standard ODE solver with $g(\bar z; p)$, where the constants $p$ have been obtained during the training. The evolved abundances in the physical space $\bar x(t)$ are then recovered by applying the decoder $\psi$ to $\bar z(t)$; cfr.~\fig{fig:recap}. In other words, one should now be able to infer the temporal evolution of $\bar x(t)$ without solving the full, relatively expensive system of ODEs.

\section{Results}\label{sect:results}

For our training data, we use the chemical network \texttt{osu\_09\_2008}\footnote{The author of the Ohio State University (OSU) chemical network is E.~Herbst. The original file is no longer reachable as of March 2021, however the code and the data employed in this paper can be found at \url{https://bitbucket.org/tgrassi/latent_ode_paper/}, commit \texttt{83300a1}.}, which includes 29~distinct species (H, H$^+$, H$_2$, H$_2^+$, H$_3^+$, O, O$^+$, OH$^+$, OH, O$_2$, O$_2^+$, H$_2$O, H$_2$O$^+$, H$_3$O$^+$, C, C$^+$, CH, CH$^+$, CH$_2$, CH$_2^+$, CH$_3$, CH$_3^+$, CH$_4$, CH$_4^+$, CH$_5^+$, CO, CO$^+$, HCO$^+$, He, He$^+$, plus electrons; \citealt{Rollig2007}) and 224~reactions. We use a constant temperature $T=50$\,K, total density $n_{\rm tot}=10^4$\,cm$^{-3}$, cosmic ray ionization rate $\zeta=10^{-16}$\,s$^{-1}$, and randomized initial conditions for C, C$^+$, O, and electrons. We assume the initial state is almost entirely molecular, i.e., $n_{\rm H_2}=n_{\rm tot}$, while $n_{\rm C}$ and $n_{\rm C^+}$ are randomly initialized in logarithmic space between $10^{-6}n_{\rm tot}$ to $10^{-3}n_{\rm tot}$, $n_{\rm O}=n_{\rm C}+n_{\rm C^+}$, and electrons are initialized on the basis of total charge neutrality. The other species are initialized to $10^{-20}n_{\rm tot}$. Note that even if these initial conditions do not represent a specific astrophysical system (e.g.~the metallicity is not, in general, constant between models), this will not affect our findings. Furthermore, because the chemical data generation and the training stage are based on randomized initial conditions, results could slightly differ between trainings.

\begin{figure}
 \includegraphics[width=\linewidth]{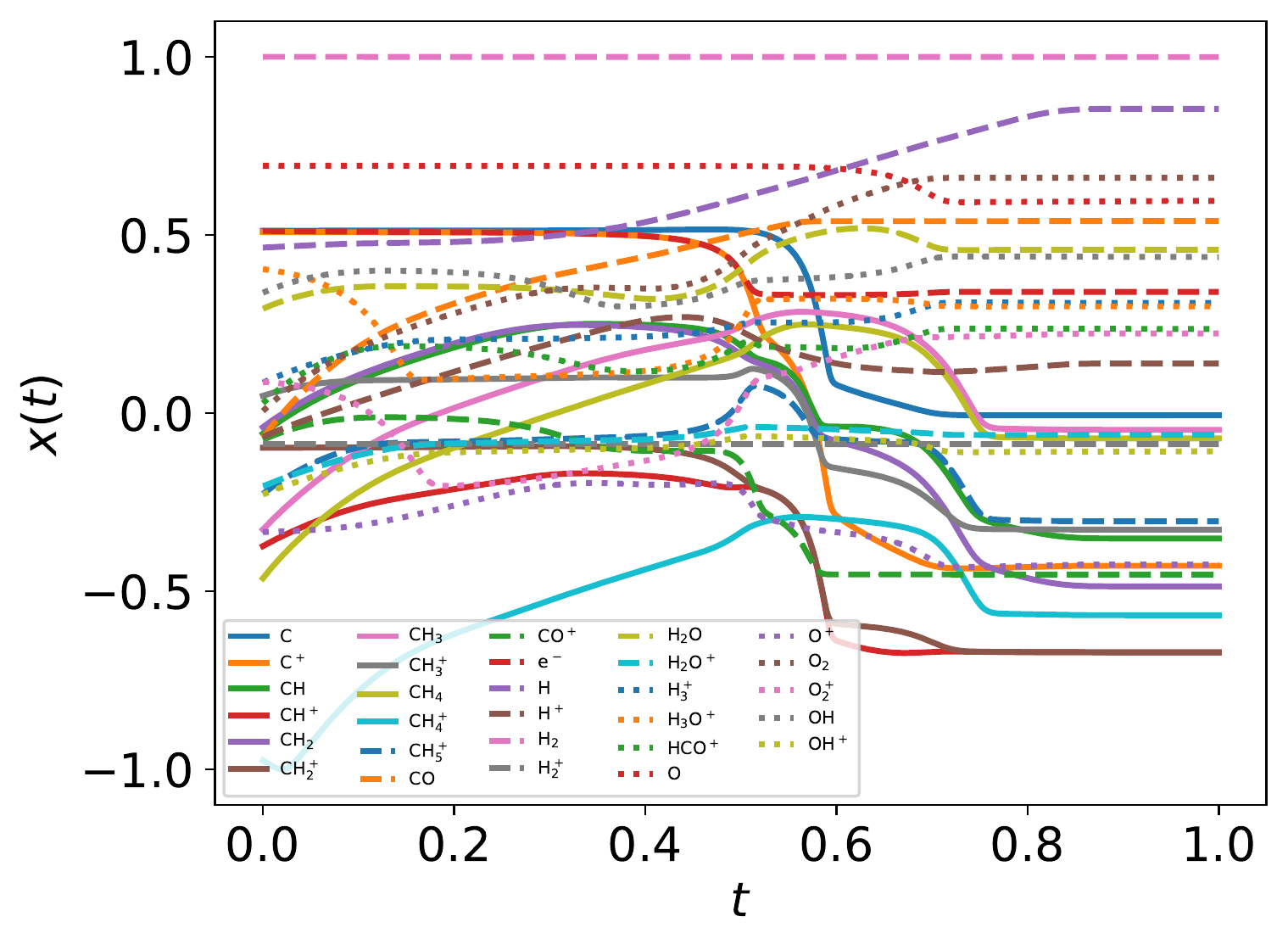}
 \caption{An example of chemical evolution in the physical space obtained by integrating $f(\bar x; k)$ in time. Each line represents the evolution of each chemical abundance $x(t)$ in time. Note that $x = \log_{10}(n/{\rm cm^{-3}})$ and normalized in the range $[-1, 1]$, while time $t = \log_{10}(time/\mathrm{yr})$ and normalized in the range $[0,1]$. The temporal grid consists of $100$ uniformly-spaced points (in logarithmic time).}\label{fig:chemistry}
\end{figure}

We integrate the ODEs with the backward differentiation formula [BDF] method of \textsc{solve\_ivp} in \textsc{scipy} \citep{Shampine1997,SciPy2020}, which provides a good compromise between computational efficiency and ease of implementation. The system of ODEs was modified (see \appx{appx:logsolver}) to solve the chemical abundances in logarithmic space, i.e., $\bar x(t) = \log_{10} \bar n(t)$, and then normalized in the range $[-1,1]$. The time evolution of $\bar x(t)$ is also computed logarithmically at 100~pre-determined grid points over an interval of $10^8$\,yr for a given randomized set of initial conditions, and then normalized in the range $[0, 1]$. A representative example of one result set is shown in~\fig{fig:chemistry}. The conversion to log-space is necessary to account for the orders of magnitudes differences spanned by the chemical abundances, while the normalization of $\bar x$ is because the activation function of the output layer is $tanh$, hence limited in the range $[-1,1]$. Time is meanwhile normalized to $[0, 1]$ for the sake of clarity.

The autoencoder consists of an encoder with dense layers of size 32, 16, and 8~nodes with ReLU activation \citep{Agarap2018}, a latent space layer of 5~nodes (i.e.~$\bar z$) and $tanh$ activation, and a decoder symmetric to the encoder, but with $tanh$ activation on the last output layer (see \appx{appx:model_layout}). The number of hidden layers and nodes employed was obtained by testing several configurations that, after approximately $10^2$ training epochs, demonstrate a good reconstruction of $\bar x(t)$, as retrieved by decoding $\bar z(t)$ including the integration of $g(\bar z; p)$ with the ODE solver.

With respect to $g(\bar z; p)$, the latent ODE system, this branch of the network consists of a custom layer with $M=5$ input nodes (i.e.~$\bar z$) and 5~output nodes (i.e.~$\dot{\bar z}$), 12~trainable parameters $p$ (one for each latent chemical reaction), and linear activation. The number of latent variables has been obtained by testing several configurations. In particular, we note that the autoencoder (i.e.~loss term $L_0$ alone) is capable of compressing the system with only two latent variables, but when we include the additional branch (i.e.~adding loss $L_1$ and $L_2$) we need at least 3~variables. If we want to add a latent chemical network that includes mass conservation (i.e.~including $L_3$), the minimum number of variables is~5. This is also the setup that is better at reproducing $\bar x(t)$ via decoding the evolving $\bar z(t)$ (see \appx{appx:latent_network}).

We employ \textsc{Keras} \citep{Keras2015} and \textsc{TensorFlow} \citep{TF2015} to build our deep neural network\footnote{\textsc{TensorFlow}~2.3.1 with CUDA~10.1, on a NVIDIA Tesla V100S; compute capability~7.0. The training wall-clock time is approximately 2.5~hours.}. We use 550~chemical models with random initial conditions, assigning 500~to the training set, and 50~to the test set. We note that by using this procedure, the independence of the test set from the training set is not guaranteed, and that the resulting latent differential equations will not work for \emph{any} possible set of initial conditions (i.e.~the latent differential equations produced by the training procedure cannot be considered to be ``universal''). That said, the final goal of such a methodology is to obtain a reduced function $g$ capable of representing the original ODE system $f$ when applied to \emph{any} $\bar x$. However, this is not the case for the results discussed in this work (see also \sect{sect:limitations}).

\begin{figure}
 \includegraphics[width=\linewidth]{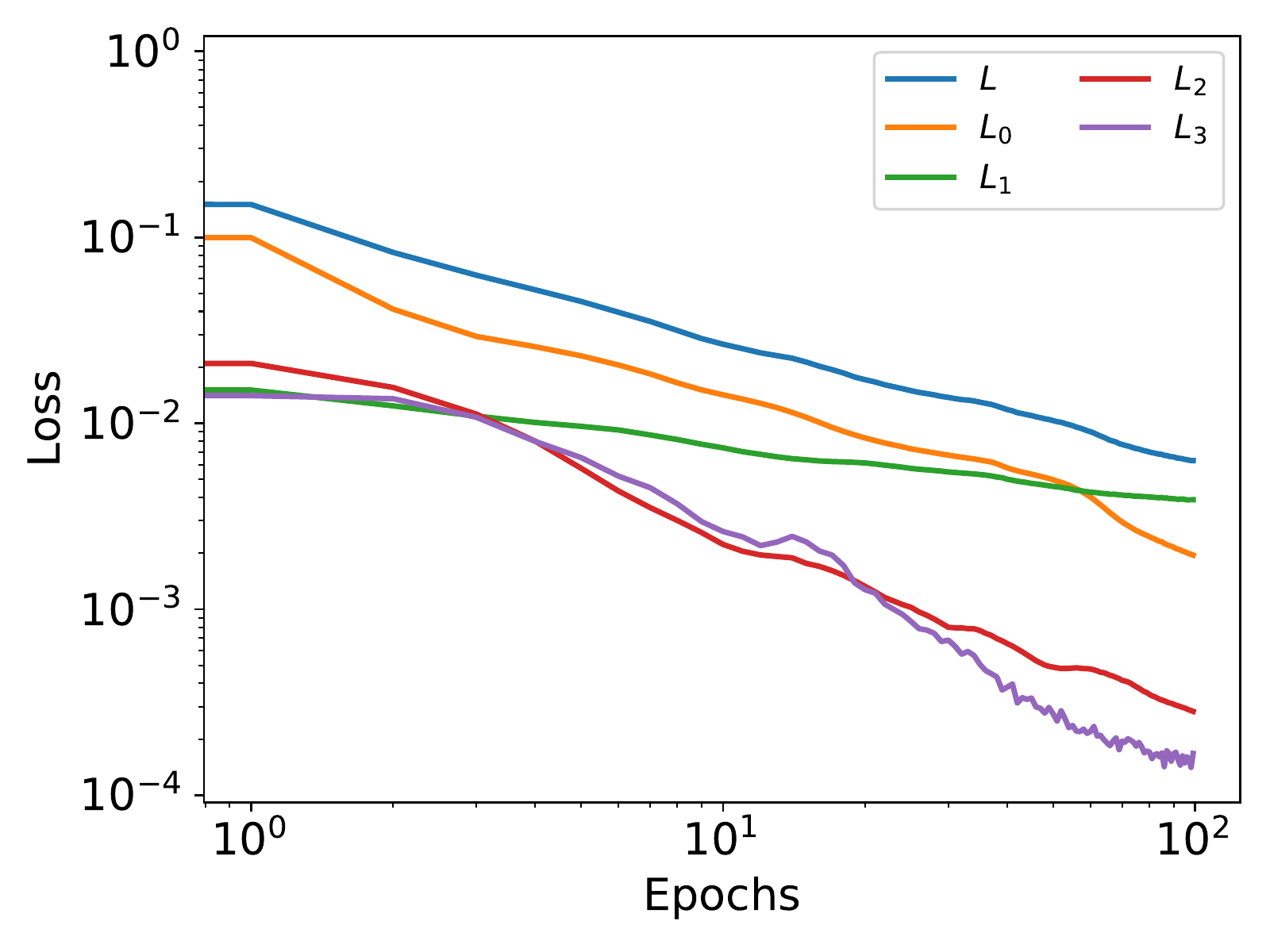}
 \caption{Loss as a function of the training epochs. The total loss $L$ (blue) is the sum of the autoencoder reconstruction loss $L_0$ (orange), the loss $L_1$ on the reconstruction of $\dot{\bar z}$ (green), the analogous $L_2$ for $\dot{\bar x}$ (red), and the mass conservation loss $L_3$ (purple). The different loss terms are defined in \sect{sect:autoencoder_plus}.}\label{fig:loss}
\end{figure}

The loss weights are $\lambda_1=\lambda_3=10^{-3}$ and $\lambda_2=10^{-2}$. We train the network using batch-size~32 and the ADAM optimizer \citep{Kingma2014} with the \textsc{Keras} default arguments, notwithstanding the learning rate, which we set to $10^{-4}$. The resulting loss as a function of training epoch for a representative training run is shown in \fig{fig:loss}. The implementation of our loss function, \eqn{eqn:loss_tot}, is described in \appx{appx:loss}. Since the training stage is quite sensitive to the initial conditions and the loss weights, we stop the iterations via visual inspection, i.e., when the reconstruction of $\bar x(t)$ is satisfactory. We plan, however, to improve this in the future with an automatic stopping criterion. The total loss ($L$, blue), is initially dominated by the autoencoder reconstruction of $\bar x$ ($L_0$, orange line in \fig{fig:loss}), and then by the $\dot{\bar z}$ reconstruction ($L_1$, green). The $\dot{\bar x}$ reconstruction ($L_2$, red) and the mass conservation loss ($L_3$ purple) are always sub-dominant, but still contribute measurably to the total loss. We note that, although $L_2$, and the other loss terms, not exhibit any strange behaviour (i.e.~continue declining), generally above $10^2$ epochs the reconstruction of $\dot{\bar z}$ begins to diverge or oscillate around the correct solution, and for this reason we interrupt the training at this point. This is probably related to the specific analytic expression use for $g$ (see the discussion in \sect{sect:limitations}).

\begin{figure}
 \includegraphics[width=\linewidth]{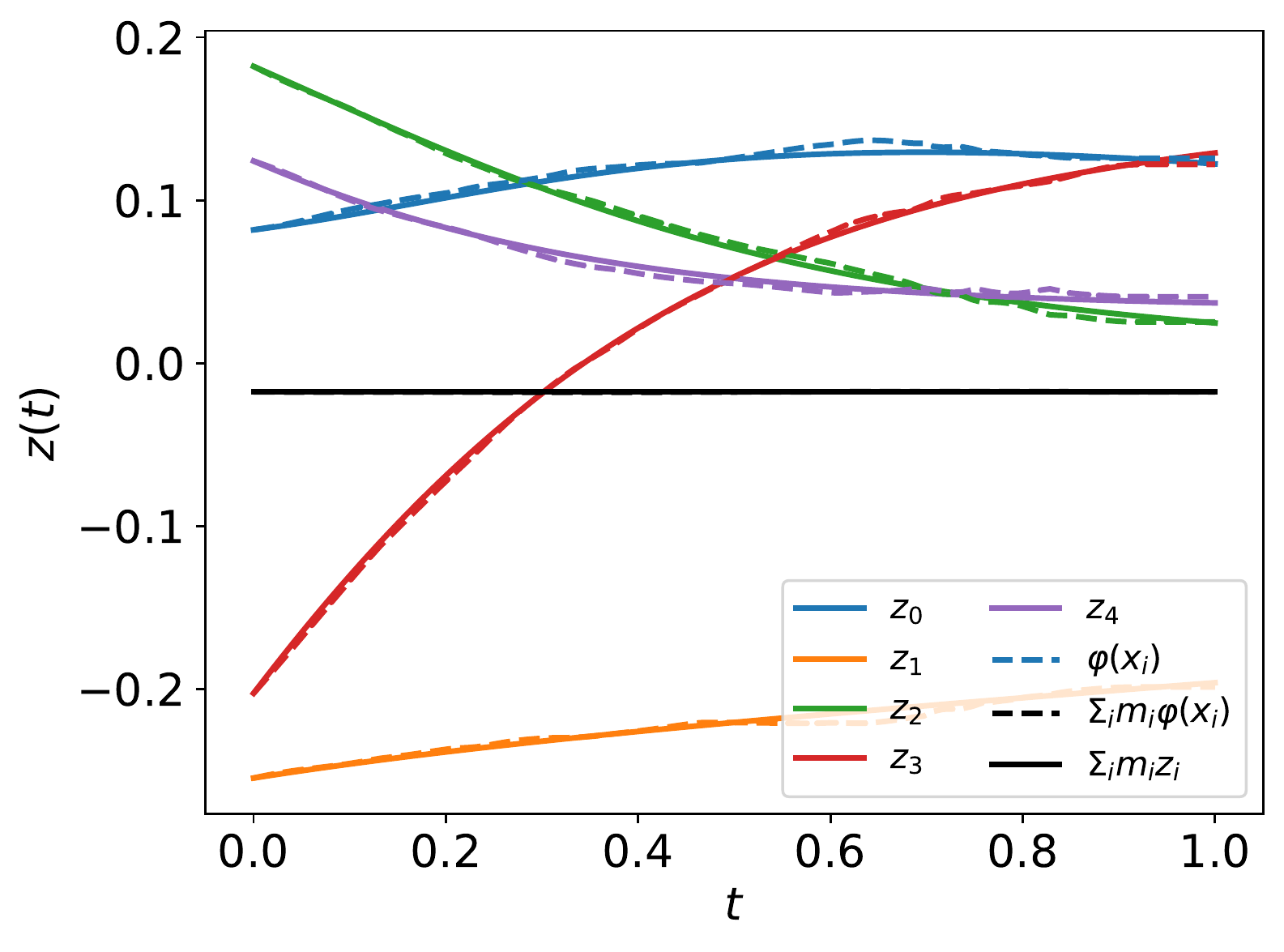}
 \caption{Evolution in time of $\bar z$ obtained by directly encoding the original data, i.e., $\varphi(\bar x)$ (dashed lines), and the corresponding evolution obtained by integrating the ODE system $g$ in the latent space starting from $\bar z(t=0)$ (solid lines). The different colors represent the 5~latent space variables (the latent ``abundances''). The black lines represent the total ``mass'' in the two cases. }\label{fig:zevol}
\end{figure}

The time evolution of $\bar z$ in the latent space for one randomly selected model of the test set is shown in \fig{fig:zevol}. The latent variables ``abundances'' are obtained from the encoded data $\varphi(\bar x(t))$ (dashed lines), while the evolution of the latent chemical network, starting from $\bar z(t=0) = \varphi(\bar x(t=0))$, is obtained using a standard ODE solver (\textsc{solve\_ivp}) to integrate $g$ in time (solid lines). This comparison reveals that it is indeed possible to evolve the chemistry in a latent compressed space and obtain results that are very close to those obtained in the physical space.  When dashed and solid lines overlap we have a perfect reconstruction. However, the difference we find in the reconstruction of $\bar z(t)$ depends again on the analytical form of $g$ that directly determines the evolution in time of the latent variables (solid lines). Recall that, in the physical space, the evolution in time of the 29~species $\bar x(t)$ is accomplished using an ODE solver that integrates $f(\bar x; k)$ built from 224~reactions, while in the latent space, we evolve in time the 5~species $\bar z(t)$ using the same ODE solver to integrate $g(\bar z; p)$ determined by 12~reactions only (cfr.~\fig{fig:method}).

\begin{figure}
 \includegraphics[width=\linewidth]{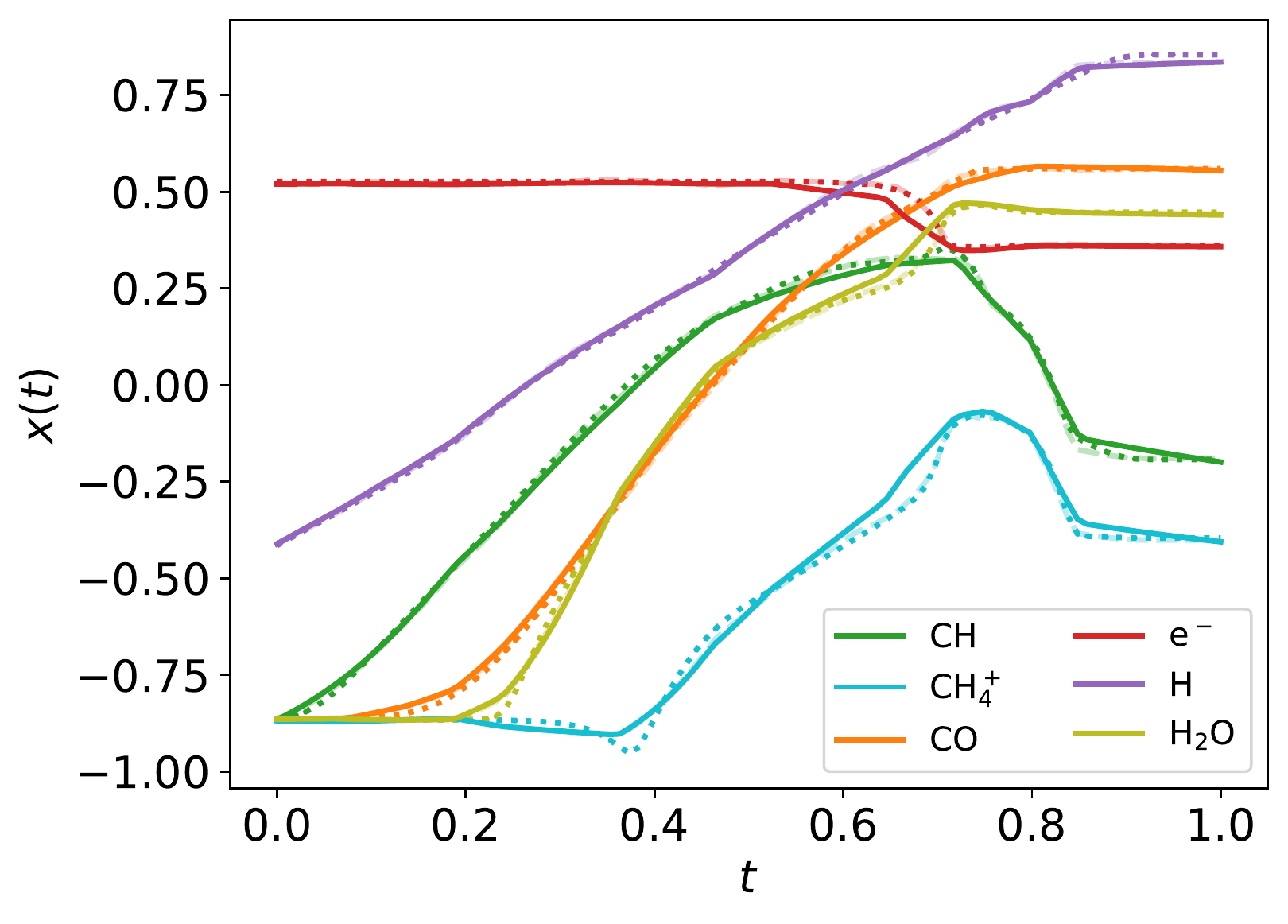}
 \caption[default]{Evolution in time of $\bar x$ for some selected species obtained by integrating the ODE in the physical space $f(\bar x; k)$ (dotted lines, ground truth), by directly autoencoding the original data, i.e., $\psi(\varphi(\bar x))$ (dashed lines), and by decoding the evolution of $\bar z$ derived by integrating the ODE system $g$ in the latent space (solid lines). This corresponds to $\psi(\bar z(t))$, where $\bar z(t)$ is represented by the solid lines in \fig{fig:zevol}). In this plot we report 6~species out of the 29 total.}\label{fig:xevol}
\end{figure}

This method should be capable of not only retrieving the temporal evolution of $\varphi(\bar x(t))$, but also of the chemical species $\bar x(t)$. In \fig{fig:xevol}, we compare for 6~selected species (out of~29) the evolution in time of the original data $\bar x(t)$ (dotted lines), the reconstructed $\bar x(t)' = \psi(\varphi(\bar x(t)))$ values (dashed), and the corresponding decoded evolution in the latent space $\psi(\bar z(t))$ (solid), where $\bar z(t)$ is evolved with the ODE solver integrating $g$. The direct autoencoder reconstruction of the original data $\psi(\varphi(\bar x))$, i.e., without integrating the differential equations in the latent space, is more accurate (as can be inferred from comparing $L_0$ and $L_1$ in \fig{fig:loss}), but this is because it is obtained directly from $\bar x(t)$, the product of the integration in physical space.
Conversely, the latent space method $\psi(\bar z)$ achieves nearly the same results (solid lines), but by evolving only the set of 5~variables $\bar z$ instead of the original $\bar x(t)$ with 29~species.
Where there are noticeable differences, we find that the largest discrepancies in \fig{fig:xevol} reflect the differences between the solid ($\bar z(t)$) and the dashed ($\varphi(\bar x(t))$) lines in \fig{fig:zevol}.

When our results are ``denormalized'' back to the original physical space of the chemical abundances (i.e.~physical units, cm$^{-3}$), the maximum relative error obtained in the test set is within an order of magnitude depending on the chemical species, but we also note that the poor feature reconstruction around $t\simeq0.6$ in \fig{fig:xevol}, play a crucial role in increasing this error.
To determine the error statistics we fit the logarithms of the relative errors $\Delta r_{i, j} = \Delta n_{i,j}\,n_{i,j}^{-1}$ with a probability density function $\mathcal{D}_i$, where $\Delta n_{i,j}$ is the difference in ``denormalized'' physical units (i.e.~cm$^{-3}$) between the actual and the predicted value of the \ith{} species abundance in the \jth{} observation (i.e.~the time-steps at which $n_{i}$ is evaluated during the test stage). Such distributions resemble a Gaussian and we therefore determine their means and standard deviations by fitting a single Gaussian profile to the distribution of the logarithm of the relative errors for each species. If we exclude H$_2$ for which we have almost perfect reconstruction, the means of the logarithms of the relative errors, $\langle\log(\Delta r)\rangle$, vary between $\sim$-2.96 and $\sim0.044$, while the standard deviations, $\sigma[\log(\Delta r)]$, vary between $\sim0.31$ and $\sim0.8$ (detailed information is reported in \appx{sect:error}). Therefore, despite the promising results (i.e.~the main features of the curves are well reproduced), the size of the error makes our method currently ineffective in replacing a standard chemical solver. Moreover, the error in the reconstructed abundances relative to the test set strong depend on the initialization of the weights of the autoencoder, as well as on the loss weights employed, which suggest that the convergence of the final parameters ($W_{ij}$ and $p$) that minimize the total loss, could be improved by a better designed $g$.

The average time\footnote{Tests performed on a single core of an Intel$^\circledR$ Core$^{\rm TM}$ i7-10750H at 2.60GHz.} to integrate $f$ during the preparation of the training set is $\sim6.5$~s wall-clock time using the BDF solver in \textsc{solve\_ivp}, while to integrate the trained, reduced ODE system $g$ we need only $\sim0.1$~s (but also using the BDF solver). In both cases the Jacobian is calculated with a finite-difference approximation. We plan to perform additional benchmark comparisons in a future work employing larger and more complicated chemical networks, and using a more efficient solver as, e.g., \textsc{dlsodes} \citep{Hindmarsh2005}.

Our results show that it is possible to obtain a considerably compressed chemical network that can be integrated with a standard ODE solver, hence reducing the computational time. However, this is not the only advantage, as the compressed solution is interpretable as a chemical network, a result that is impossible to achieve by only analyzing the weights of the hidden layers of a standard deep neural network. Hence, the evolution is more controllable and expressible in chemical terms, and could potentially expose some relevant information about the original chemical network, as well as allow analytical integration in time in some special cases.

\section{Limitations and Outlook}\label{sect:limitations}

This paper presents a novel method to solving time-dependent chemistry, but this particular study should be considered as a proof-of-concept and has some limitations that we discuss below. We plan to address these shortcomings in the near future before deploying the method to more realistic and more challenging scientific applications, e.g., the coupled calculation of the (thermo)chemical evolution of a large (magneto)hydrodynamic simulation.

First, as with many data-driven or machine-learning based methods, the training set should be considerably large to enable the method to sample the complete spectrum of possible inputs and outputs. However, it is important to remark that, in contrast to other deep learning methods, our approach has the advantage that the chemical network in the latent space is transparent and interpretable. Nonetheless, the number of data points required to train our neural network is $2N\times N_{\rm \Delta t}\times N_{\rm models}$, where $N$ is the number of chemical species, $N_{\rm \Delta t}$ the number of time-steps during the evolution when the chemical data is recorded, and $N_{\rm models}$ the number of models with different initial conditions; the factor of $2$ indicates that both $\bar x$ and $\dot{\bar x}$ (the time derivatives) are needed for the training set. While the order of magnitude of $N$ and $N_{\rm tsteps}$ is likely to vary less, $N_{\rm models}$ might vary considerably depending on the problem in order to cover the parameter space \citep{deMijolla2019}. The size of the training data set also places limitations on how large a parameter space can be explored. Although the results presented in \sect{sect:results} have a small memory footprint ($<1$~GB), and hence can be obtained with an average consumer-level GPU card, some set ups, which will be the subject of a forthcoming work, required up to $30$\,GB of memory, handled in our case by a \textsc{NVIDIA Tesla V100S} GPU with $32$\,GB of memory.

Another challenge is the definition of an analytical expression for $g$. This is a critical aspect, since this is where the interpretability of the method arises. In our application, we consider the simplest case of only two latent ``atoms'' (see \appx{appx:latent_network}). The resultant latent space consists of only 5~variables (or latent ``abundances''), and for these reasons the maximum number of possible latent reactions is limited to~12. However, larger chemical networks in the physical space might only be compressible with $>5$ latent variables, and consequently the increased number of possible variable combinations might produce a relatively large latent chemical network, reducing or nullifying the advantage of the low-dimensionality latent chemical space. That said, how the minimum number of required latent variables changes with the size and complexity of a chemical network is currently unknown. As previously discussed, to cope with this problem, a promising method is to employ the concept of parsimonious representation \citep{Champion2019,Champion2019b}. In this approach, an additional loss term is added to reduce the number of non-zero coefficients $p$ in $g(\bar z; p)$ to the minimum required set. As before, it is worth mentioning that, at the moment, the design and understanding of chemical networks in the latent space is uncharted territory.

\begin{figure}
 \includegraphics[width=\linewidth]{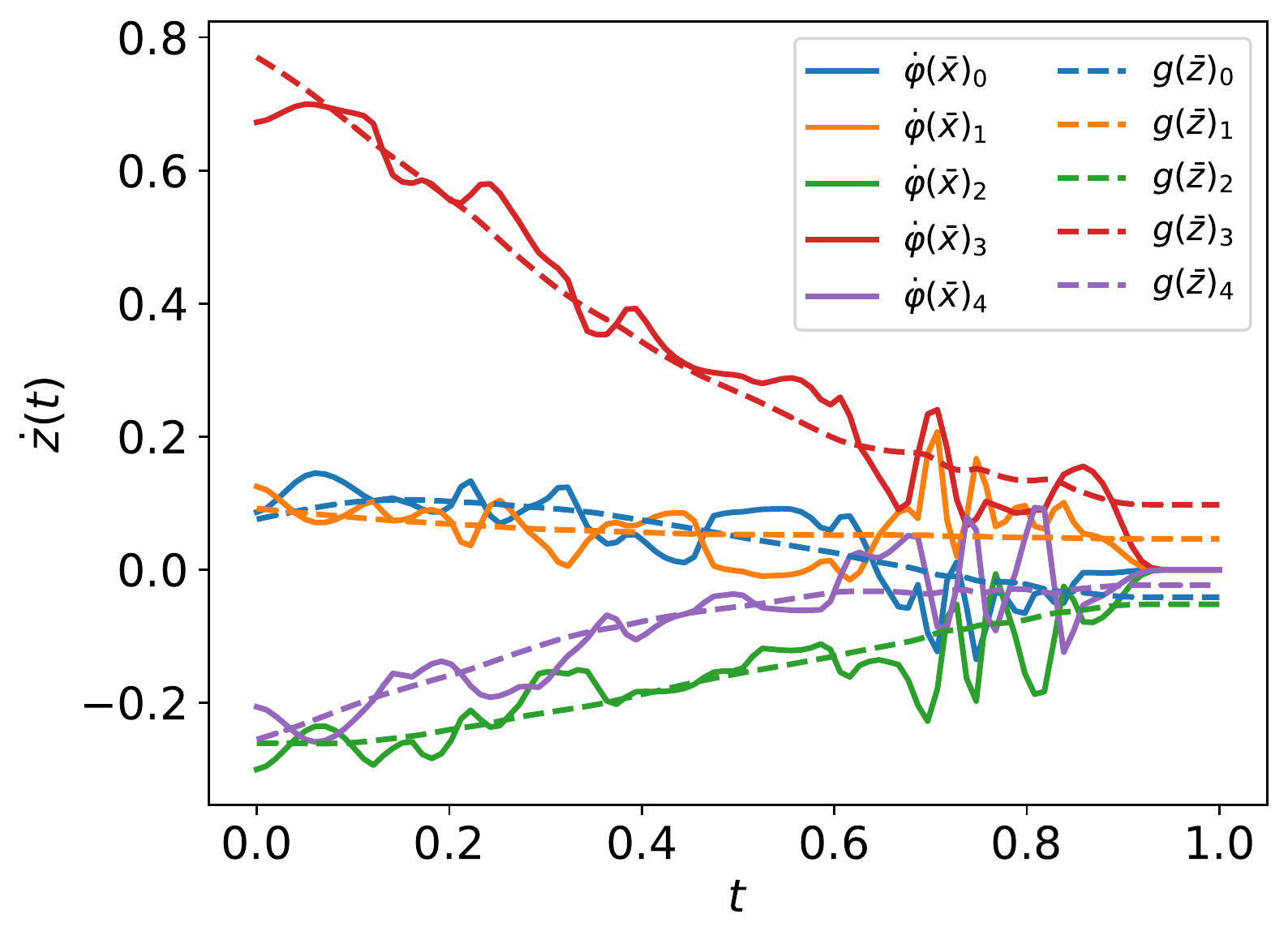}
 \caption[default]{Time derivatives of the encoded data, $\dot{\varphi}(\bar{x})$ (solid lines), compared to the latent ODEs, $g(\bar{z})$ (dashed), both as a function of time. The different colors denote the different components of the latent space.}\label{fig:zgrad}
\end{figure}

Furthermore, to achieve an accurate reconstruction of the time evolution of $\bar x$, the latent ODE $g$ should be capable of reproducing the high-frequency behaviour of the encoded data, $\phi(\bar{x}_i$). In \fig{fig:zgrad} we report the comparison of the time derivatives of the five components of the latent space (i.e.~$\dot{\bar z}$) computed via the trained function $g(\bar z; p)$ (dashed) and via the gradient in time of the encoded $\bar x$ (solid). With a perfect reconstruction the two representations of $\dot{\bar z}$ should overlap, but in our case the derivative computed from $g$ resembles a time average of the derivative of the encoded $\bar x$, suggesting that with our current implementation the fine, high-frequency behavior in the latent space is poorly reconstructed.

In terms of technical issues, the success of the training depends on the user-defined loss weights $\lambda_i$ in \eqn{eqn:loss_tot}. If, for example, $L_0$ dominates over the other losses (see~\fig{fig:loss}), the autoencoder perfectly reproduces $\bar x$ as $\psi(\varphi(\bar x))$, but fails in reproducing $\bar x$ from $\psi(\bar z)$ when $\bar z$ is evolved using $g$. Conversely, if $L_1$ or $L_2$ dominates the total loss, the autoencoder fails completely to reproduce the original data, and no useful solution can be found. Additionally, the absolute value of the loss is defined with respect to an error, i.e.~$|L|<\varepsilon$. For example in $L_0$, \eqn{eqn:loss0}, the loss is calculated as the sum of the difference between the absolute values of the original and reconstructed abundances, squared. However, since the abundances have been normalized (\sect{sect:results}), the contribution of each chemical species to $L_0$ is the same, which may not be appropriate given the often orders of magnitude difference between abundances of different species. Depending on the aims of the specific astrophysical problem, this particular loss term can be replaced by a weighted loss, where the user defines which species play a key role in $L_0$. Another approach which may prove beneficial in some contexts is to use the difference between the relative abundances of the species, rather than the absolute one. Analogous considerations apply to $L_1$, $L_2$, and $L_3$.

Concerning the present study, the adopted chemical network and the resulting data have a limited amount of variability. Not in terms of the evolution of the chemical abundances (see \fig{fig:chemistry}), but rather in terms of the limited initial conditions. In fact, a constant temperature and cosmic-ray ionization rate translates into time-independent reaction rate coefficients $k$, and thus the same holds true for the corresponding coefficients $p$ in the latent space. The evolution of the temperature is a key aspect in many astrophysical models, and hence it should be included as an additional variable and evolved alongside the chemical abundances. Indeed, since our current model produces an interpretable representation in the latent space, adding the temperature might lead to a system of latent ODEs $g$ that gives additional insights on the interplay between chemistry and thermal processes. This will be addressed in future work.

A key aspect and the final goal of our method is to obtain a ``universal'' set of differential equations in the latent space ($g$) that, within a given approximation error, effectively replaces the analogous function in the physical space ($f$). If such a system exists, in principle \emph{any} trajectory $\bar x(t)$ produced by applying $f(\bar x)$ has its own corresponding trajectory $\bar z(t)$ that can be advanced in time by applying $g(\bar z)$. If this achievement is obtained, any arbitrarily uncorrelated test set of models can be reproduced by our trained framework, as far as this test set is produced by using $f$. Within this context, in this paper we are not obtaining such a result because (i) the functions $g$ is not the latent representation of $f$, and/or (ii) the autoencoder is not a sufficiently accurate compressed representation of the physical space.

In (magneto)hydrodynamic simulations, the (thermo)chemical evolution is often included using an operator splitting technique, i.e., alternating between (thermo)chemical evolution and dynamics (but with a communally determined time step). In the current implementation of our method, this implies that $\bar x(t)$ needs to be encoded to $\bar z(t)$, evolved with $g$ in the latent space to $z(t+\Delta t)$, and decoded again to $\bar x(t+\Delta t)$ during \emph{each dynamical time step}. This could result in a significant cost overhead if the computational time saved by solving $g$ instead of $f$ is smaller than the time spent applying $\varphi$ and $\psi$.  In the present test, the time spent for one encode/decode is negligible ($\lesssim0.008$~s) when compared to integrating $f$ directly ($\sim6.5$s). However, avoiding the need to encode/decode each time step when coupling this method to a dynamical simulation will be addressed in a future work.

In simulations that solve (thermo)chemistry and dynamics together via operator splitting, care must be taken to minimize the propagation of errors in the chemical abundances over time due to the advection of species. This can be dealt with using so-called ``consistent multi-fluid advection'' schemes \citep{Plewa1999, Glover2010, Grassi2017} that ensure conservation of, e.g.~the metallicity. Analogously, it might be possible within the current method to include additional losses designed to conserve elemental abundances (in our case, H, C, and~O)

The method presented in this study shows promising results, and represents a novel approach to the problem of reducing the computational impact of modeling (thermo)chemical evolution, particularly in the context of large scale dynamical simulations. However, several limitations present in the current implementation suggest that further exploration and a deeper analysis of the methodology are required. It is clear however that machine learning is a rapidly and continually growing field, and that faster and more capable hardware becomes available at regular intervals. As such, we are confident that the class of methods to which this study belongs will prove capable at efficiently reducing the computational impact of not only time-dependent chemical evolution, but for other systems of astrophysically-relevant differential equations.

\section{Conclusions}\label{sect:conclusions}

In this work, we describe the theoretical foundations and a first application of a novel data-driven method aimed at using autoencoders to reduce the complexity of multi-dimensional and time-dependent chemistry, and to reproduce the temporal evolution of the chemistry when coupled to a learned, latent system of ODEs.\\

\noindent In summary, we find that:
\begin{itemize}[noitemsep,topsep=1pt]
 \item This approach can reduce the number of chemical species (i.e.~dimensions) by encoding the physical space into a low-dimensional latent space.

 \item This compression not only manages to preserve the information stored in the original data, it also ensures that the evolution of the compressed variables is representable by another set of ordinary differential equations corresponding to a latent ``chemical network'' with a considerably smaller number of reactions, and that the time-dependent chemical abundances can subsequently be accurately reconstructed.

 \item In the proof-of-concept application presented in this work, we are capable of reducing a chemical network with 224~reactions and 29~species into a compressed network with 12~reactions and 5~species that can be evolved forward in time with a standard ODE solver.

 \item Integrating a considerably smaller chemical network in time permits a considerable computational speed-up relative to integrating the original network. In our preliminary tests we obtained a $\times65$ speed-up.

 \item The interpretability of the latent variables and ODEs provides an advantage  compared to opaque standard machine learning methods of dimensionality reduction. Moreover, interpretability could allow us to better understand the intrinsic chemical properties of a network in the physical-space through identification of specific characteristics or behaviors in the latent space that might otherwise be hidden.
\end{itemize}
$\phantom{a}$

\noindent We also note that the current implementation includes a set of limitations that needs to be solved in the future:
\begin{itemize}[noitemsep,topsep=1pt]
 \item Similar to most of the data-driven machine learning methods, the training dataset needs to be considerably large to completely explore the full spectrum of variability. Additionally, to achieve a successful training, the different loss terms need to be tuned by hand in order to obtain a quick and effective convergence.

 \item The topology of the compressed chemical network needs to be outlined by the user (i.e.~$g$ currently needs to be defined analytically), which is not always trivial. However, various techniques have been proposed and will be explored in a future work to simplify this task.

 \item This method has been tested in a rather controlled environment (e.g.~constant temperature, density, and cosmic-ray ionization rate) with a limited amount of variability in the initial conditions, and will need to be generalized in the future.
\end{itemize}
$\phantom{a}$

In conclusion, despite the limitations and some technical issues that we will address in future works, the method presented here is a promising new approach to solving the problem of complicated chemical evolution and its often high computational cost. The code is publicly-available at \url{https://bitbucket.org/tgrassi/latent_ode_paper/}. Finally, it is important to note that the chemical data generation and the training stage in this work are both based on randomized initial conditions, hence code output and the results in this paper could differ slightly.

\section*{Acknowledgments}
Wt thank the referee for the useful comments that improved the quality of the paper.
We thank T.~Hoffmann for his help in installing and configuring the hardware employed in this paper.
SB is financially supported by ICM (Iniciativa Cient\'ifica Milenio) via N\'ucleo Milenio en Tecnolog\'ia e Investigaci\'on Transversal para explorar Agujeros Negros Supermasivos (\#NCN19\_058), and BASAL Centro de Astrofisica y Tecnologias Afines (CATA) AFB-17002.
BE acknowledges support from the DFG cluster of excellence ``Origin and Structure of the Universe''
(\verb+http://www.universe-cluster.de/+).
GP acknowledges support from the DFG Research Unit ``Transition Disks’' (FOR~2634/1, ER~685/8-1).
JPR acknowledges support from the Virginia Initiative on Cosmic Origins (VICO), the National Science Foundation (NSF) under grant nos.\ AST-1910106 and AST-1910675, and NASA via the Astrophysics Theory Program under grant no.\ 80NSSC20K0533.
This work was funded by the DFG Research Unit FOR 2634/1 ER685/11-1.
This research was supported by the Excellence Cluster ORIGINS, which is funded by the Deutsche Forschungsgemeinschaft (DFG, German Research Foundation) under Germany's Excellence Strategy - EXC-2094 - 390783311.

\section*{Software}
This work made use of the following open source projects: \textsc{Matplotlib} \citep{Matplotlib2007}, \textsc{SciPy} \citep{SciPy2020}, \textsc{NumPy} \citep{NumPy2020}, \textsc{Inkscape} (\url{https://inkscape.org}), \textsc{Keras} \citep{Keras2015}, and \textsc{TensorFlow} \citep{TF2015}.




\bibliographystyle{aa}
\bibliography{mybib} 

\begin{thebibliography}{80}
\expandafter\ifx\csname natexlab\endcsname\relax\def\natexlab#1{#1}\fi

\bibitem[{Abadi {et~al.}(2015)Abadi, Agarwal, Barham, Brevdo, Chen, Citro,
  Corrado, Davis, Dean, Devin, Ghemawat, Goodfellow, Harp, Irving, Isard, Jia,
  Jozefowicz, Kaiser, Kudlur, Levenberg, Man\'{e}, Monga, Moore, Murray, Olah,
  Schuster, Shlens, Steiner, Sutskever, Talwar, Tucker, Vanhoucke, Vasudevan,
  Vi\'{e}gas, Vinyals, Warden, Wattenberg, Wicke, Yu, \& Zheng}]{TF2015}
Abadi, M., Agarwal, A., Barham, P., {et~al.} 2015, {TensorFlow}: Large-Scale
  Machine Learning on Heterogeneous Systems, software available from
  tensorflow.org

\bibitem[{{Agarap}(2018)}]{Agarap2018}
{Agarap}, A.~F. 2018, arXiv e-prints, arXiv:1803.08375

\bibitem[{{Akimkin} {et~al.}(2013){Akimkin}, {Zhukovska}, {Wiebe}, {Semenov},
  {Pavlyuchenkov}, {Vasyunin}, {Birnstiel}, \& {Henning}}]{Akimkin2013}
{Akimkin}, V., {Zhukovska}, S., {Wiebe}, D., {et~al.} 2013, \apj, 766, 8

\bibitem[{{Bai}(2016)}]{Bai2016}
{Bai}, X.-N. 2016, \apj, 821, 80

\bibitem[{Baulch {et~al.}(2005)Baulch, Bowman, Cobos, Cox, Just, Kerr, Pilling,
  Stocker, Troe, Tsang, Walker, \& Warnatz}]{Baulch2005}
Baulch, D.~L., Bowman, C.~T., Cobos, C.~J., {et~al.} 2005, J. Phys. Chem. Ref.
  Data, 34, 757

\bibitem[{{Bovino} {et~al.}(2019){Bovino}, {Ferrada-Chamorro}, {Lupi},
  {Sabatini}, {Giannetti}, \& {Schleicher}}]{Bovino2019}
{Bovino}, S., {Ferrada-Chamorro}, S., {Lupi}, A., {et~al.} 2019, \apj, 887, 224

\bibitem[{{Bovino} {et~al.}(2013){Bovino}, {Grassi}, {Latif}, \&
  {Schleicher}}]{Bovino2013}
{Bovino}, S., {Grassi}, T., {Latif}, M.~A., \& {Schleicher}, D.~R.~G. 2013,
  \mnras, 434, L36

\bibitem[{{Bruderer} {et~al.}(2009){Bruderer}, {Doty}, \&
  {Benz}}]{Bruderer2009}
{Bruderer}, S., {Doty}, S.~D., \& {Benz}, A.~O. 2009, \apjs, 183, 179

\bibitem[{Brunton {et~al.}(2016)Brunton, Proctor, \& Kutz}]{Brunton2016}
Brunton, S.~L., Proctor, J.~L., \& Kutz, J.~N. 2016, Proceedings of the
  National Academy of Sciences, 113, 3932

\bibitem[{{Chakraborty} {et~al.}(2017){Chakraborty}, {Tomsett}, {Raghavendra},
  {Harborne}, {Alzantot}, {Cerutti}, {Srivastava}, {Preece}, {Julier}, {Rao},
  {Kelley}, {Braines}, {Sensoy}, {Willis}, \& {Gurram}}]{Chakraborty2017}
{Chakraborty}, S., {Tomsett}, R., {Raghavendra}, R., {et~al.} 2017, in 2017
  IEEE SmartWorld, Ubiquitous Intelligence Computing, Advanced Trusted
  Computed, Scalable Computing Communications, Cloud Big Data Computing,
  Internet of People and Smart City Innovation
  (SmartWorld/SCALCOM/UIC/ATC/CBDCom/IOP/SCI), 1--6

\bibitem[{Champion {et~al.}(2019)Champion, Lusch, Kutz, \&
  Brunton}]{Champion2019}
Champion, K., Lusch, B., Kutz, J.~N., \& Brunton, S.~L. 2019, Proceedings of
  the National Academy of Sciences, 116, 22445

\bibitem[{{Champion} {et~al.}(2020){Champion}, {Zheng}, {Aravkin}, {Brunton},
  \& {Kutz}}]{Champion2019b}
{Champion}, K., {Zheng}, P., {Aravkin}, A.~Y., {Brunton}, S.~L., \& {Kutz},
  J.~N. 2020, IEEE Access, 8, 169259

\bibitem[{{Chen} {et~al.}(2018){Chen}, {Rubanova}, {Bettencourt}, \&
  {Duvenaud}}]{Chen2018}
{Chen}, R. T.~Q., {Rubanova}, Y., {Bettencourt}, J., \& {Duvenaud}, D. 2018,
  arXiv e-prints, arXiv:1806.07366

\bibitem[{Chollet {et~al.}(2015)}]{Keras2015}
Chollet, F. {et~al.} 2015, Keras, \url{https://keras.io}

\bibitem[{Choudhary {et~al.}(2020)Choudhary, Lindner, Holliday, Miller, Sinha,
  \& Ditto}]{Choudhary2020}
Choudhary, A., Lindner, J.~F., Holliday, E.~G., {et~al.} 2020, Phys. Rev. E,
  101, 062207

\bibitem[{Curtis {et~al.}(2017)Curtis, Niemeyer, \& Sung}]{Curtis2017}
Curtis, N.~J., Niemeyer, K.~E., \& Sung, C.-J. 2017, Combustion and Flame, 179,
  312

\bibitem[{{de Mijolla} {et~al.}(2019){de Mijolla}, {Viti}, {Holdship},
  {Manolopoulou}, \& {Yates}}]{deMijolla2019}
{de Mijolla}, D., {Viti}, S., {Holdship}, J., {Manolopoulou}, I., \& {Yates},
  J. 2019, \aap, 630, A117

\bibitem[{Duff {et~al.}(1986)Duff, Duff, Erisman, Reid, \& Reid}]{Duff1986}
Duff, N., Duff, I., Erisman, A., Reid, C., \& Reid, J. 1986, Direct Methods for
  Sparse Matrices, Monographs on numerical analysis (Clarendon Press)

\bibitem[{{Garrod}(2008)}]{Garrod2008}
{Garrod}, R.~T. 2008, \aap, 491, 239

\bibitem[{{Glover} \& {Clark}(2012)}]{Glover2012}
{Glover}, S. C.~O. \& {Clark}, P.~C. 2012, \mnras, 421, 116

\bibitem[{{Glover} {et~al.}(2010){Glover}, {Federrath}, {Mac Low}, \&
  {Klessen}}]{Glover2010}
{Glover}, S.~C.~O., {Federrath}, C., {Mac Low}, M.~M., \& {Klessen}, R.~S.
  2010, \mnras, 404, 2

\bibitem[{Gondara(2016)}]{Gondara2016}
Gondara, L. 2016, 2016 IEEE 16th International Conference on Data Mining
  Workshops (ICDMW)

\bibitem[{{Gong} {et~al.}(2017){Gong}, {Ostriker}, \& {Wolfire}}]{Gong2017}
{Gong}, M., {Ostriker}, E.~C., \& {Wolfire}, M.~G. 2017, \apj, 843, 38

\bibitem[{{Grassi} {et~al.}(2012){Grassi}, {Bovino}, {Gianturco}, {Baiocchi},
  \& {Merlin}}]{Grassi2012}
{Grassi}, T., {Bovino}, S., {Gianturco}, F.~A., {Baiocchi}, P., \& {Merlin}, E.
  2012, \mnras, 425, 1332

\bibitem[{{Grassi} {et~al.}(2017){Grassi}, {Bovino}, {Haugb{\o}lle}, \&
  {Schleicher}}]{Grassi2017}
{Grassi}, T., {Bovino}, S., {Haugb{\o}lle}, T., \& {Schleicher}, D.~R.~G. 2017,
  \mnras, 466, 1259

\bibitem[{{Grassi} {et~al.}(2013){Grassi}, {Bovino}, {Schleicher}, \&
  {Gianturco}}]{Grassi2013}
{Grassi}, T., {Bovino}, S., {Schleicher}, D., \& {Gianturco}, F.~A. 2013,
  \mnras, 431, 1659

\bibitem[{{Grassi} {et~al.}(2011){Grassi}, {Merlin}, {Piovan}, {Buonomo}, \&
  {Chiosi}}]{Grassi2011}
{Grassi}, T., {Merlin}, E., {Piovan}, L., {Buonomo}, U., \& {Chiosi}, C. 2011,
  arXiv e-prints, arXiv:1103.0509

\bibitem[{{Grassi} {et~al.}(2019){Grassi}, {Padovani}, {Ramsey}, {Galli},
  {Vaytet}, {Ercolano}, \& {Haugb{\o}lle}}]{Grassi2019}
{Grassi}, T., {Padovani}, M., {Ramsey}, J.~P., {et~al.} 2019, \mnras, 484, 161

\bibitem[{{Gressel} {et~al.}(2020){Gressel}, {Ramsey}, {Brinch}, {Nelson},
  {Turner}, \& {Bruderer}}]{Gressel2020}
{Gressel}, O., {Ramsey}, J.~P., {Brinch}, C., {et~al.} 2020, \apj, 896, 126

\bibitem[{Harris {et~al.}(2020)Harris, Millman, van~der Walt, Gommers,
  Virtanen, Cournapeau, Wieser, Taylor, Berg, Smith, Kern, Picus, Hoyer, van
  Kerkwijk, Brett, Haldane, Fernández~del Río, Wiebe, Peterson,
  Gérard-Marchant, Sheppard, Reddy, Weckesser, Abbasi, Gohlke, \&
  Oliphant}]{NumPy2020}
Harris, C.~R., Millman, K.~J., van~der Walt, S.~J., {et~al.} 2020, Nature, 585,
  357–362

\bibitem[{{Henning} \& {Semenov}(2013)}]{Henning2013}
{Henning}, T. \& {Semenov}, D. 2013, Chemical Reviews, 113, 9016

\bibitem[{{Herbst} \& {van Dishoeck}(2009)}]{Herbst2009}
{Herbst}, E. \& {van Dishoeck}, E.~F. 2009, \araa, 47, 427

\bibitem[{{Heyl} {et~al.}(2020){Heyl}, {Viti}, {Holdship}, \&
  {Feeney}}]{Heyl2020}
{Heyl}, J., {Viti}, S., {Holdship}, J., \& {Feeney}, S.~M. 2020, arXiv
  e-prints, arXiv:2010.02877

\bibitem[{Hindmarsh {et~al.}(2005)Hindmarsh, Brown, Grant, Lee, Serban,
  Shumaker, \& Woodward}]{Hindmarsh2005}
Hindmarsh, A.~C., Brown, P.~N., Grant, K.~E., {et~al.} 2005, ACM Trans. Math.
  Softw., 31, 363–396

\bibitem[{Hoffmann {et~al.}(2019)Hoffmann, Fröhner, \& Noé}]{Hoffmann2019}
Hoffmann, M., Fröhner, C., \& Noé, F. 2019, The Journal of Chemical Physics,
  150, 025101

\bibitem[{{Holdship} {et~al.}(2018){Holdship}, {Jeffrey}, {Makrymallis},
  {Viti}, \& {Yates}}]{Holdship2018}
{Holdship}, J., {Jeffrey}, N., {Makrymallis}, A., {Viti}, S., \& {Yates}, J.
  2018, \apj, 866, 116

\bibitem[{Hunter(2007)}]{Matplotlib2007}
Hunter, J.~D. 2007, Computing in Science \& Engineering, 9, 90

\bibitem[{{Ilee} {et~al.}(2017){Ilee}, {Forgan}, {Evans}, {Hall}, {Booth},
  {Clarke}, {Rice}, {Boley}, {Caselli}, {Hartquist}, \& {Rawlings}}]{Ilee2017}
{Ilee}, J.~D., {Forgan}, D.~H., {Evans}, M.~G., {et~al.} 2017, \mnras, 472, 189

\bibitem[{{Jolliffe}(2002)}]{Jolliffe2002}
{Jolliffe}, I. 2002, Principal component analysis (New York: Springer Verlag)

\bibitem[{Jørgensen {et~al.}(2020)Jørgensen, Belloche, \&
  Garrod}]{Jorgensen2020}
Jørgensen, J.~K., Belloche, A., \& Garrod, R.~T. 2020, Annual Review of
  Astronomy and Astrophysics, 58, 727

\bibitem[{{Kingma} \& {Ba}(2014)}]{Kingma2014}
{Kingma}, D.~P. \& {Ba}, J. 2014, arXiv e-prints, arXiv:1412.6980

\bibitem[{{Kingma} \& {Welling}(2013)}]{Kingma2013}
{Kingma}, D.~P. \& {Welling}, M. 2013, arXiv e-prints, arXiv:1312.6114

\bibitem[{Kramer(1991)}]{Kramer1991}
Kramer, M.~A. 1991, AIChE Journal, 37, 233

\bibitem[{{Lecun} {et~al.}(1998){Lecun}, {Bottou}, {Bengio}, \&
  {Haffner}}]{Lecun1998}
{Lecun}, Y., {Bottou}, L., {Bengio}, Y., \& {Haffner}, P. 1998, Proceedings of
  the IEEE, 86, 2278

\bibitem[{{Lipton}(2016)}]{Lipton2016}
{Lipton}, Z.~C. 2016, arXiv e-prints, arXiv:1606.03490

\bibitem[{{Long} {et~al.}(2017){Long}, {Lu}, {Ma}, \& {Dong}}]{Long2017}
{Long}, Z., {Lu}, Y., {Ma}, X., \& {Dong}, B. 2017, arXiv e-prints,
  arXiv:1710.09668

\bibitem[{{Lupi} \& {Bovino}(2020)}]{Lupi2020}
{Lupi}, A. \& {Bovino}, S. 2020, \mnras, 492, 2818

\bibitem[{{McGuire}(2018)}]{McGuire2018}
{McGuire}, B.~A. 2018, \apjs, 239, 17

\bibitem[{{Miller}(2017)}]{Miller2017}
{Miller}, T. 2017, arXiv e-prints, arXiv:1706.07269

\bibitem[{Nejad(2005)}]{Nejad2005}
Nejad, L. 2005, Astrophysics and Space Science, 299, 1

\bibitem[{{Nicolini} \& {Frezzato}(2013)}]{Nicolini2013}
{Nicolini}, P. \& {Frezzato}, D. 2013, \jcp, 138, 234102

\bibitem[{Perini {et~al.}(2012)Perini, Galligani, \& Reitz}]{Perini2012}
Perini, F., Galligani, E., \& Reitz, R.~D. 2012, Energy \& Fuels, 26, 4804

\bibitem[{{Plewa} \& {M{\"u}ller}(1999)}]{Plewa1999}
{Plewa}, T. \& {M{\"u}ller}, E. 1999, \aap, 342, 179

\bibitem[{{Rab} {et~al.}(2017){Rab}, {Elbakyan}, {Vorobyov}, {G{\"u}del},
  {Dionatos}, {Audard}, {Kamp}, {Thi}, {Woitke}, \& {Postel}}]{Rab2017}
{Rab}, C., {Elbakyan}, V., {Vorobyov}, E., {et~al.} 2017, \aap, 604, A15

\bibitem[{Rackauckas {et~al.}(2019)Rackauckas, Innes, Ma, Bettencourt, White,
  \& Dixit}]{Rackauckas2019}
Rackauckas, C., Innes, M., Ma, Y., {et~al.} 2019, CoRR, abs/1902.02376
  [\eprint[arXiv]{1902.02376}]

\bibitem[{Raissi \& Karniadakis(2018)}]{Raissi2018}
Raissi, M. \& Karniadakis, G.~E. 2018, Journal of Computational Physics, 357,
  125

\bibitem[{Raissi {et~al.}(2019)Raissi, Perdikaris, \& Karniadakis}]{Raissi2019}
Raissi, M., Perdikaris, P., \& Karniadakis, G. 2019, Journal of Computational
  Physics, 378, 686

\bibitem[{{R{\"o}llig} {et~al.}(2007){R{\"o}llig}, {Abel}, {Bell}, {Bensch},
  {Black}, {Ferland}, {Jonkheid}, {Kamp}, {Kaufman}, {Le Bourlot}, {Le Petit},
  {Meijerink}, {Morata}, {Ossenkopf}, {Roueff}, {Shaw}, {Spaans}, {Sternberg},
  {Stutzki}, {Thi}, {van Dishoeck}, {van Hoof}, {Viti}, \&
  {Wolfire}}]{Rollig2007}
{R{\"o}llig}, M., {Abel}, N.~P., {Bell}, T., {et~al.} 2007, \aap, 467, 187

\bibitem[{Rubanova {et~al.}(2019)Rubanova, Chen, \& Duvenaud}]{Rubanova2019}
Rubanova, Y., Chen, R. T.~Q., \& Duvenaud, D.~K. 2019, in Advances in Neural
  Information Processing Systems 32, ed. H.~Wallach, H.~Larochelle,
  A.~Beygelzimer, F.~d'~Alch\'{e}-Buc, E.~Fox, \& R.~Garnett (Curran
  Associates, Inc.), 5320--5330

\bibitem[{{Ruffle} {et~al.}(2002){Ruffle}, {Rae}, {Pilling}, {Hartquist}, \&
  {Herbst}}]{Ruffle2002}
{Ruffle}, D.~P., {Rae}, J.~G.~L., {Pilling}, M.~J., {Hartquist}, T.~W., \&
  {Herbst}, E. 2002, \aap, 381, L13

\bibitem[{Rumelhart {et~al.}(1988)Rumelhart, Hinton, \&
  Williams}]{Rumelhart1988}
Rumelhart, D.~E., Hinton, G.~E., \& Williams, R.~J. 1988, Learning
  Representations by Back-Propagating Errors (Cambridge, MA, USA: MIT Press),
  696–699

\bibitem[{{Semenov} {et~al.}(2010){Semenov}, {Hersant}, {Wakelam}, {Dutrey},
  {Chapillon}, {Guilloteau}, {Henning}, {Launhardt}, {Pi{\'e}tu}, \&
  {Schreyer}}]{Semenov2010}
{Semenov}, D., {Hersant}, F., {Wakelam}, V., {et~al.} 2010, \aap, 522, A42

\bibitem[{{Semenov} {et~al.}(2004){Semenov}, {Wiebe}, \&
  {Henning}}]{Semenov2004}
{Semenov}, D., {Wiebe}, D., \& {Henning}, T. 2004, \aap, 417, 93

\bibitem[{Shampine \& Reichelt(1997)}]{Shampine1997}
Shampine, L.~F. \& Reichelt, M.~W. 1997, SIAM J. Sci. Comput., 18, 1–22

\bibitem[{{Shlens}(2014)}]{Shlens2014}
{Shlens}, J. 2014, arXiv e-prints, arXiv:1404.1100

\bibitem[{{Sipil{\"a}} {et~al.}(2010){Sipil{\"a}}, {Hugo}, {Harju}, {Asvany},
  {Juvela}, \& {Schlemmer}}]{Sipila2010}
{Sipil{\"a}}, O., {Hugo}, E., {Harju}, J., {et~al.} 2010, \aap, 509, A98

\bibitem[{{Tian} {et~al.}(2013){Tian}, {Saito}, {Preis}, {Garcia}, {Kozhukhov},
  {Masten}, {Cherkasov}, \& {Panchenko}}]{Tian2013}
{Tian}, X., {Saito}, H., {Preis}, S.~V., {et~al.} 2013, in 2013 IEEE
  International Symposium on Parallel Distributed Processing, Workshops and Phd
  Forum, 1149--1158

\bibitem[{Tibshirani(1996)}]{Tibshirani1996}
Tibshirani, R. 1996, Journal of the Royal Statistical Society. Series B
  (Methodological), 58, 267

\bibitem[{Tupper(2002)}]{Tupper2002}
Tupper, P. 2002, BIT Numerical Mathematics, 42, 447

\bibitem[{{Valorani} \& {Goussis}(2001)}]{Valorani2001}
{Valorani}, M. \& {Goussis}, D.~A. 2001, Journal of Computational Physics, 169,
  44

\bibitem[{Virtanen {et~al.}(2020)Virtanen, Gommers, Oliphant, Haberland, Reddy,
  Cournapeau, Burovski, Peterson, Weckesser, Bright, {van der Walt}, Brett,
  Wilson, Millman, Mayorov, Nelson, Jones, Kern, Larson, Carey, Polat, Feng,
  Moore, {VanderPlas}, Laxalde, Perktold, Cimrman, Henriksen, Quintero, Harris,
  Archibald, Ribeiro, Pedregosa, {van Mulbregt}, \& {SciPy 1.0
  Contributors}}]{SciPy2020}
Virtanen, P., Gommers, R., Oliphant, T.~E., {et~al.} 2020, Nature Methods, 17,
  261

\bibitem[{{Wakelam} {et~al.}(2012){Wakelam}, {Herbst}, {Loison}, {Smith},
  {Chandrasekaran}, {Pavone}, {Adams}, {Bacchus-Montabonel}, {Bergeat},
  {B{\'e}roff}, {Bierbaum}, {Chabot}, {Dalgarno}, {van Dishoeck}, {Faure},
  {Geppert}, {Gerlich}, {Galli}, {H{\'e}brard}, {Hersant}, {Hickson},
  {Honvault}, {Klippenstein}, {Le Picard}, {Nyman}, {Pernot}, {Schlemmer},
  {Selsis}, {Sims}, {Talbi}, {Tennyson}, {Troe}, {Wester}, \&
  {Wiesenfeld}}]{Wakelam2012}
{Wakelam}, V., {Herbst}, E., {Loison}, J.~C., {et~al.} 2012, \apjs, 199, 21

\bibitem[{{Walsh} {et~al.}(2014){Walsh}, {Millar}, {Nomura}, {Herbst}, {Widicus
  Weaver}, {Aikawa}, {Laas}, \& {Vasyunin}}]{Walsh2014}
{Walsh}, C., {Millar}, T.~J., {Nomura}, H., {et~al.} 2014, \aap, 563, A33

\bibitem[{{Wiebe} {et~al.}(2003){Wiebe}, {Semenov}, \& {Henning}}]{Wiebe2003}
{Wiebe}, D., {Semenov}, D., \& {Henning}, T. 2003, \aap, 399, 197

\bibitem[{Wiewel {et~al.}(2019)Wiewel, Becher, \& Thuerey}]{Wiewel2018}
Wiewel, S., Becher, M., \& Thuerey, N. 2019, Computer Graphics Forum, 38, 71

\bibitem[{{Woitke} {et~al.}(2009){Woitke}, {Kamp}, \& {Thi}}]{Woitke2009}
{Woitke}, P., {Kamp}, I., \& {Thi}, W.~F. 2009, \aap, 501, 383

\bibitem[{{Xu} {et~al.}(2019){Xu}, {Bai}, {{\"O}berg}, \& {Zhang}}]{Xu2019}
{Xu}, R., {Bai}, X.-N., {{\"O}berg}, K., \& {Zhang}, H. 2019, \apj, 872, 107

\bibitem[{{Y{\i}ld{\i}z} {et~al.}(2019){Y{\i}ld{\i}z}, {Heinonen}, \&
  {L{\"a}hdesm{\"a}ki}}]{Yildiz2019}
{Y{\i}ld{\i}z}, {\c{C}}., {Heinonen}, M., \& {L{\"a}hdesm{\"a}ki}, H. 2019,
  arXiv e-prints, arXiv:1905.10994

\bibitem[{{Yoon} \& {Kwak}(2018)}]{Yoon2018}
{Yoon}, J. \& {Kwak}, K. 2018, in Journal of Physics Conference Series, Vol.
  1031, Journal of Physics Conference Series, 012023

\bibitem[{Zhou \& Paffenroth(2017)}]{Zhou2017}
Zhou, C. \& Paffenroth, R.~C. 2017, in Proceedings of the 23rd ACM SIGKDD
  International Conference on Knowledge Discovery and Data Mining, 665--674

\end{thebibliography}



\begin{appendix}

\section{Relation between the number of species and the number of reactions}\label{sect:relation}
\begin{figure}
 \includegraphics[width=\linewidth]{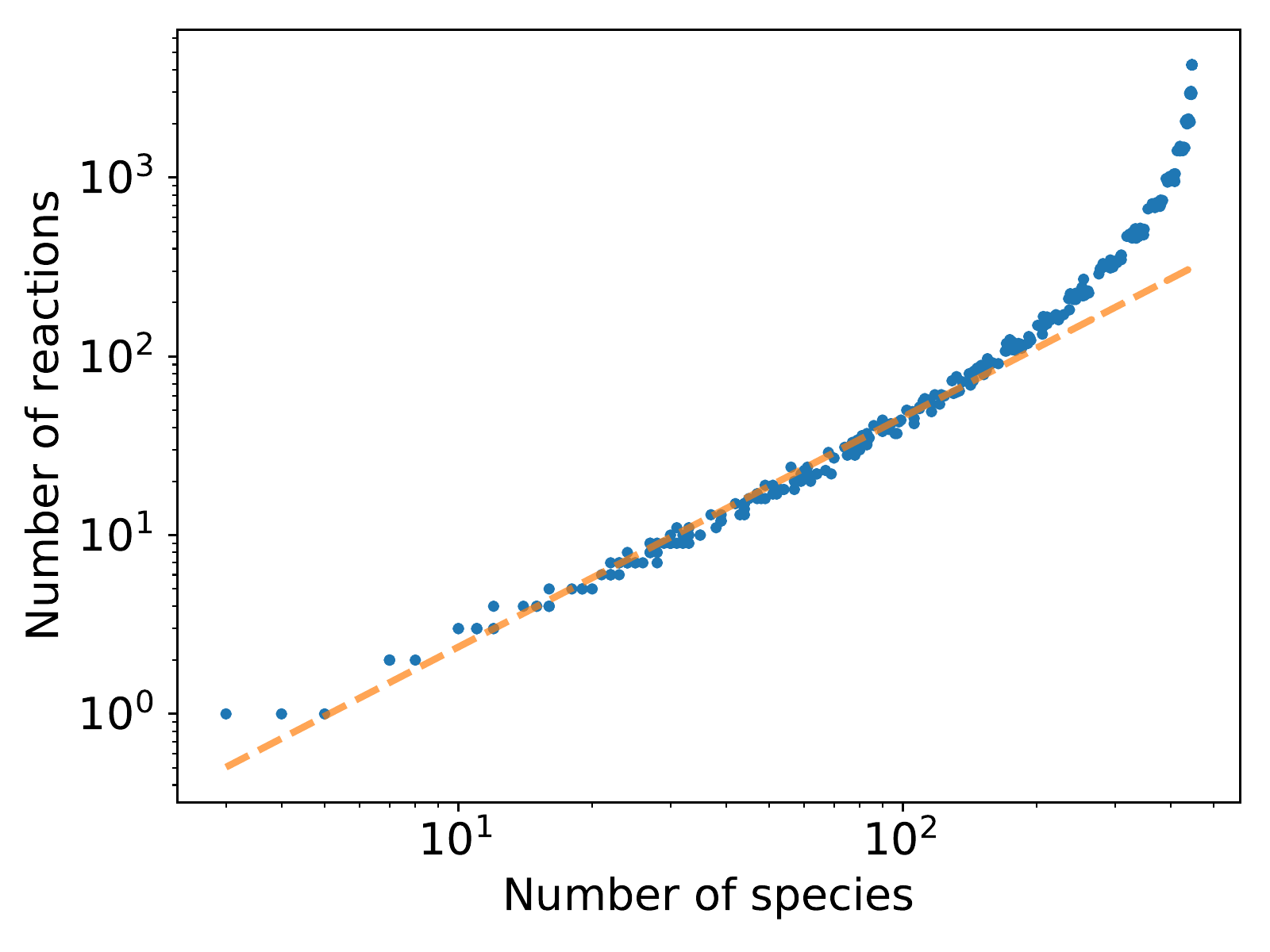}
 \caption{Number of chemical reactions as a function of the number of species, constructed by randomly removing blocks of 400 reactions. The dashed line is a power-law in $N$ with an exponent of 1.3.}\label{fig:relation}
\end{figure}

We tested the relation between the number of reactions (${\cal N}_{\rm R}$) and the number of species ($N$) for our ``complete'' \texttt{osu\_09\_2008} network (4457~reactions) by randomly removing reactions and counting the number of species left. The results are reported in \fig{fig:relation} where we randomly remove blocks of 400 reactions, and show a ${\cal N}_{\rm  R} \propto N^{1.3}$ power-law relation. The increasing slope at the rightmost side (i.e.~ at a large number of species) of the plot is because several species are found in a many different reactions, but only drop out when all of those reactions have been removed. Note that this assumes our complete chemical network is roughly representative of the true overall catalog of chemical reactivity of the included species. Note also that the computational cost has a more complicated relation with the number of reactions and/or species, since the cost also depends on the different time-scales in the network (related to stiffness) and the sparsity of the Jacobian, which depends on the types of reactions involved (see \sect{sect:ode}).

\section{The latent chemical network and its analytical representation}\label{appx:latent_network}

One of the most complex tasks of this method is to define the analytical form for $\dot{\bar z} = g(\bar z; p)$. At the present stage, we do not have any automatic (machine learning-based or otherwise) method to find the optimal analytical form of $g$, not even from its parameters $p$ and some additional loss term, as for example a parsimonious loss term, $||\Xi||_1$ \citep{Champion2019}, and mass conservation. The approach we follow then is to create a small network that presents mass conservation and includes two-body reactions for non-linearity. The simplest approach is to consider only two ``atoms'', A and B, to create a limited number of ``molecules'', namely A, B, AA, BB, and AB. These are the 5 latent variables/species $\bar z$. With mass conservation and assuming that A and B have the same mass, so that $m_{\rm A}= m_{\rm B}=1$ and $m_{\rm AA}= m_{\rm BB}=m_{\rm AB}=2$, see \eqn{eqn:loss_mass}, the possible ``reactions'' are 6~forward and the corresponding 6~reverse:
\beqa
 \rm A + A &\toot& \rm AA\\
 \rm A + B &\toot& \rm AB\\
 \rm B + B &\toot& \rm BB\\
 \rm AA + BB &\toot& \rm AB + AB\\
 \rm AA + B &\toot& \rm AB + A\\
 \rm BB + A &\toot& \rm AB + B
\eeqa
with the following ODEs (cfr.~\eqn{eqn:ode})
\beqa\label{eqn:zdot}
 \dot z_{\rm A} &=& - 2F_0 - F_1 + F_4 - F_5\\
 \dot z_{\rm B} &=& - F_1 - 2F_2 - F_4 + F_5\\
 \dot z_{\rm AA} &=&  + F_0 - F_3 - F_4\\
 \dot z_{\rm BB} &=&  + F_2 - F_3 - F_5\\
 \dot z_{\rm AB} &=&  + F_1 + 2F_3 + F_4 + F_5
\eeqa
where $F_i = R_i - R_{i+6}$, and $R_0= p_0 z_{\rm A}^2$, $R_1 = p_1 z_{\rm A} z_{\rm B}$, \dots, $R_5 = p_5 z_{\rm BB} z_{\rm A}$ are the forward reaction rates, while $R_6= p_6 z_{\rm AA}$, $R_7 = p_7 z_{\rm AB}$, \dots, $R_{11} = p_{11} z_{\rm AB}z_{\rm B}$ are the reverse reaction rates, and where $p$ are the rate coefficients determined during training.

To implement these ODEs as $g$ in the custom layer of \textsc{TensorFlow} we define the ``tensorial'' form
\beq
  \bar R = \bar p \cdot \bar Z\,,
\eeq
where $\bar Z$ is an array that contains the reactant products, one element per reaction (e.g.~$Z_0 = z_{\rm A} z_{\rm B}$ or more generally $Z_i=R_i/p_i$). Next, we define
\beq
  \dot{\bar z} = g(\bar z; p) = S \times \bar R\,,
\eeq
where $S$ is a matrix with size equal to the number of species times the number of reactions. Values $S_{ij} > 0$ indicate that the \ith{} species is a product of the \jth{} reaction, while $S_{ij} < -1$ indicates a reactant, and $S_{ij} = 0$ indicates that the species is neither a reactant nor a product of the \jth{} reaction. There is a matrix $S$ for each ODE, e.g., for \eqn{eqn:zdot} it is
\beq S^T =
\begin{pmatrix}[r]
 -2 &  1 &  0 &  0 &  0\\
  2 & -1 &  0 &  0 &  0\\
 -1 &  0 &  1 & -1 &  0\\
  1 &  0 & -1 &  1 &  0\\
  0 &  0 &  0 & -2 &  1\\
  0 &  0 &  0 &  2 & -1\\
  0 & -1 &  2 &  0 & -1\\
  0 &  1 & -2 &  0 &  1\\
  1 & -1 &  1 & -1 &  0\\
 -1 &  1 & -1 &  1 &  0\\
 -1 &  0 &  1 &  1 & -1\\
  1 &  0 & -1 & -1 &  1\\
\end{pmatrix}\,.
\eeq

The \textsc{TensorFlow} code is automatically generated by the utility \textsc{network2tensor.py} in the project repository.

It is worth noting that, if the number of latent variables increases, this approach might produce a considerably large chemical network, diminishing the aims and power of the method. For this reason, in future work, we will improve on this aspect by developing additional constraints to better design the latent network. More details can be found in the \texttt{Gex} class in \texttt{autoencoder\_plus/test.py} in the project repository \url{https://bitbucket.org/tgrassi/latent_ode_paper/}.

\section{Log-space ODE solver}\label{appx:logsolver}

We modified the system of ODEs in order to integrate both abundances and time directly in logarithmic space. In particular, if $\tau = \log_{10}(t)$ and $y=\log_{10}(x)$ we can write $\dd x = x \ln(10)\, \dd y$ and $\dd \tau = t \ln(10)\, \dd \tau$ to obtain a new right-hand side
\beq
 \frac{\dd y}{\dd \tau} = f'(y) = \frac{10^\tau}{10^y} f(10^y)\,,
\eeq
that can be integrated with a standard BDF solver. A limitation of this method is that $t=0$ must be replaced with a non-zero value, in our case the integration starts from $t=10^{-6}$~yr. Chemical species meanwhile have a lower bound of $10^{-20}n_{\rm tot}$, where ${n_{\rm tot}=10^4\,{\rm cm}^{-3}}$.

\section{Model layout}\label{appx:model_layout}
\begin{figure}
 \includegraphics[width=\linewidth]{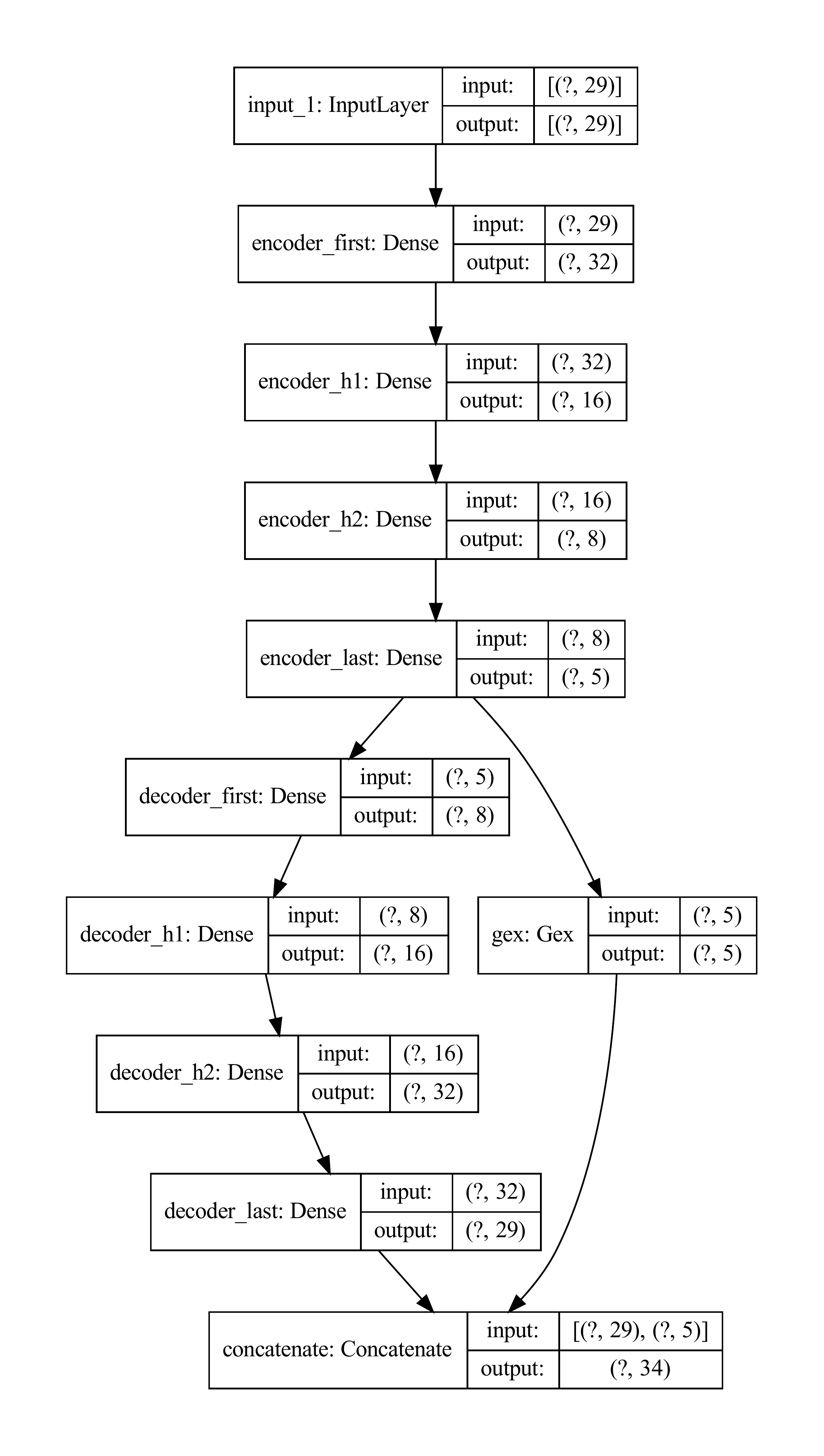}
 \caption{Model layout from the \texttt{plot\_model} utility in \textsc{Keras}. The upper part represents the encoder, the left branch the decoder, and the right the custom layer \texttt{gex}, i.e.~$g$. The output of \texttt{encoder\_last} is the latent space, i.e.~$\bar z$. \texttt{Concatenate} is a dummy layer that concatenates the output of the two branches. The question marks indicate the unknown batch size, that is defined at runtime. To be compared with \fig{fig:sketch}.}\label{fig:model}
\end{figure}

\fig{fig:model} shows the model layout as given by the \texttt{plot\_model} utility in \textsc{Keras}. The model consists of an encoder, a decoder, and the $g$ layer, here labeled \texttt{gex}. This can be compared with \fig{fig:sketch}.

The input layer (\texttt{InputLayer}) consists of 29~nodes corresponding to the chemical species $\bar x$. The 5~latent variables $\bar z$ are the output of the encoder (layer \texttt{encoder\_last}) and input for the first layer of the decoder (\texttt{decoder\_first}), aas well as the layer that represents the ODE system $g$ (\texttt{gex}). The decoder produces $\bar x'$ as output (\texttt{decoder\_last}), and hence has 29~output units, while \texttt{gex} has 5~output units, i.e.~$\dot{\bar z}$. \textsc{Keras} requires that these outputs be concatenated into a single layer with no trainable parameters and no activation function. The question marks in \fig{fig:model} indicate the batch size that is defined at run-time.

\section{Loss term implementation}\label{appx:loss}

To implement the loss terms in \textsc{Keras} and \textsc{TensorFlow}, we need to have access to $\partial_x\varphi(\bar x)$ and $\partial_z\psi(\bar z)$, i.e.~the differential variation of the encoder and decoder with respect to the variables $x$ and $z$, respectively (\sect{sect:autoencoder_plus}). To this aim, since we use the ``eager'' execution model\footnote{\url{https://www.tensorflow.org/guide/eager}} of \textsc{TensorFlow}, we take advantage of \texttt{GradientTape}, which allows us to keep track of the variation of the output of one or more layers as a function of the variation of the input quantities.

To compute the gradient, e.g.~$\partial_x\varphi(\bar x)$, instead of using the \texttt{gradient} function that automatically sums $\partial \varphi_i / \partial x_j$ along $i$, we employ \texttt{batch\_jacobian} that does not compute any sum, and allows us to write
\vspace{-2mm}
\begin{verbatim}
with tf.GradientTape() as tape:
  tape.watch(x)
  zenc = encoder(x)
dphi_dx = tape.batch_jacobian(zenc, x)
\end{verbatim}
where \texttt{encoder} is $\varphi(\bar x)$, \texttt{dphi\_dx} is the Jacobian matrix $\partial \varphi_i / \partial x_j$, and \texttt{x} is $\bar x$.
Then $\dot{\bar x}\,\partial_x\varphi(\bar x)$, i.e.~$\sum_j \dot x_j \partial \varphi_i / \partial x_j$, is implemented as \texttt{tf.linalg.matvec(dphi\_dx, xdot)}, where \texttt{tf.linalg.matvec} is the \textsc{TensorFlow} matrix multiplication operator, and \texttt{xdot} is $\dot{\bar x}$. Similar considerations apply to $\partial_z\psi(\bar z)$. More details can be found in the \texttt{loss} function in \texttt{autoencoder\_plus/test.py} in the project repository \url{https://bitbucket.org/tgrassi/latent\_ode\_paper/}.

\section{Test set error}\label{sect:error}
\begin{figure}
 \includegraphics[width=\linewidth]{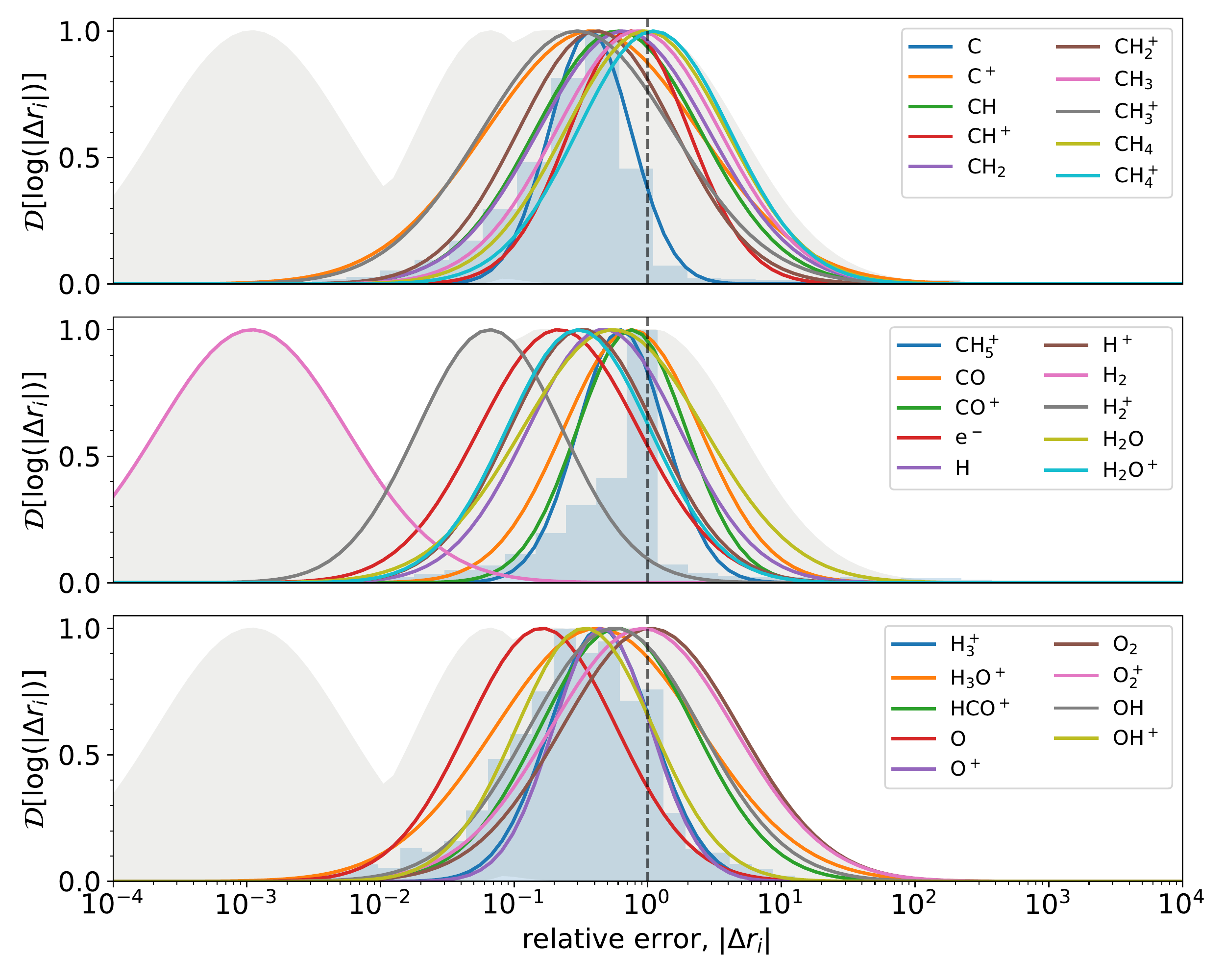}
 \caption{Gaussian fitting functions of the logarithm of the relative error distributions for each chemical species. For three selected species (C, CH$_5^+$, and H$_3^+$) we also report the normalized histogram of the error counts (light blue bars). The vertical dashed line indicates a relative error of~$1$. Note that the plotted distributions are normalized to the maximum. As a reference, the gray-shaded area is the envelope of all the curves. See also \tab{tab:error}.}\label{fig:error}
\end{figure}

Here we report on the probability density functions $\mathcal{D}_i$ and subsequent Gaussian fits to the logarithm of the relative errors of each chemical species, as discussed in \sect{sect:results}. In \tab{tab:error} we indicate the mean and the variance of the Gaussian fitting functions reported in \fig{fig:error}. Note that the values reported in \tab{tab:error} depend on the specific training history, and might differ by approximately $20\%$ between training runs.

\begin{table}
    \centering
    \begin{tabular}{llllll}
        \hline
        Species &  $\langle\log(\Delta r)\rangle$ & Std.\ Dev. & Species & $\log(\Delta r)$ & Std.\ Dev.\\
        \hline
C                    & $-0.432$ & $0.307$ & C$^+$                & $-0.415$ & $0.795$\\
CH                   & $-0.225$ & $0.619$ & CH$^+$               & $-0.139$ & $0.447$\\
CH$_2$               & $-0.180$ & $0.637$ & CH$_2^+$             & $-0.383$ & $0.585$\\
CH$_3$               & $-0.074$ & $0.611$ & CH$_3^+$             & $-0.527$ & $0.714$\\
CH$_4$               & $-0.014$ & $0.603$ & CH$_4^+$             & $+0.044$ & $0.570$\\
CH$_5^+$             & $-0.199$ & $0.328$ & CO                   & $-0.119$ & $0.513$\\
CO$^+$               & $-0.121$ & $0.408$ & e$^-$                & $-0.666$ & $0.598$\\
H                    & $-0.328$ & $0.568$ & H$^+$                & $-0.490$ & $0.544$\\
H$_2$                & $-2.962$ & $0.707$ & H$_2^+$              & $-1.181$ & $0.542$\\
H$_2$O               & $-0.257$ & $0.681$ & H$_2$O$^+$           & $-0.520$ & $0.537$\\
H$_3^+$              & $-0.351$ & $0.386$ & H$_3$O$^+$           & $-0.378$ & $0.765$\\
HCO$^+$              & $-0.228$ & $0.579$ & O                    & $-0.792$ & $0.560$\\
O$^+$                & $-0.341$ & $0.366$ & O$_2$                & $+0.022$ & $0.664$\\
O$_2^+$              & $-0.039$ & $0.685$ & OH                   & $-0.246$ & $0.640$\\
OH$^+$               & $-0.471$ & $0.512$ & & & 
    \end{tabular}
    \caption{Mean and standard deviation for the Gaussian fits to the distributions of the logarithm of the relative errors as discussed in the text. Note that it is the mean of the logarithm of the relative error, $\langle\log(\Delta r)\rangle$, and corresponding standard deviations, $\sigma[\log(\Delta r)]$, that are reported. See also \fig{fig:error}.}\label{tab:error}
\end{table}

\end{appendix}

\end{document}